\documentclass[rmp,aps,twocolumn,nofootinbib]{revtex4}

\usepackage{graphicx}
\usepackage{hyperref}

\begin{document}
\title{Order and quantum phase transitions in the cuprate superconductors}
\preprint{cond-mat/0211005}
\author{Subir Sachdev}
\email{subir.sachdev@yale.edu}
\affiliation{Department of Physics,
Yale University, P.O. Box 208120, New Haven CT 06520-8120}

\begin{abstract}
It is now widely accepted that the cuprate superconductors are
characterized by the same long-range order as that present in the
Bardeen-Cooper-Schrieffer (BCS) theory: that associated with the
condensation of Cooper pairs. The author argues that many physical
properties of the cuprates require interplay with additional order
parameters associated with a proximate Mott insulator. A
classification of Mott insulators in two dimensions is proposed.
Experimental evidence so far shows that the class appropriate to
the cuprates has collinear spin correlations, bond order, and
confinement of neutral, spin $S=1/2$ excitations. Proximity to
second-order quantum phase transitions associated with these
orders, and with the pairing order of BCS, has led to systematic
predictions for many physical properties. In this context the
author reviews the results of recent neutron scattering, fluxoid
detection, nuclear magnetic resonance, and scanning tunnelling
microscopy experiments.
\end{abstract}

\date{October 28, 2002}
\maketitle
\tableofcontents

\section{INTRODUCTION}
\label{sec:intro}

The discovery of high temperature superconductivity in the cuprate
series of compounds by \textcite{bednorz86} has strongly
influenced the development of condensed matter physics. It
stimulated a great deal of experimental work on the synthesis and
characterization of a variety of related intermetallic compounds.
It also reinvigorated theoretical study of electronic systems with
strong correlations. Technological applications of these materials
have also appeared, and could become more widespread.

Prior to this discovery, it was widely assumed that all known
superconductors, or superfluids of neutral fermions such as
$^3$He, were described by the theory of Bardeen, Cooper and
Schrieffer (BCS) \cite{bcs}. Certainly, the quantitative successes
of BCS theory in describing an impressive range of phenomena in
the lower temperature superconductors make it one of the most
successful physical theories ever proposed. Soon after the
discovery of the high temperature superconductors, it became clear
that many of their properties, and especially those at
temperatures ($T$) above the superconducting critical temperature
($T_c$), could not be quantitatively described by the BCS theory.
Overcoming this failure has been an important motivation for
theoretical work in the past decade.

One of the purposes of this article is to present an updated
assessment of the applicability of the BCS theory to the cuprate
superconductors. We will restrict our attention to physics at very
low temperature associated with the nature of the ground state and
its elementary excitations. This will allow us to focus on sharp,
qualitative distinctions. In particular, we will avoid the regime
of temperatures above $T_c$, where it is at least possible that
any failure of the BCS theory is a symptom of our inability to
make accurate quantitative predictions in a strong coupling
regime, rather than our having missed a qualitatively new type of
order. Also, while this article will present a unified view of the
important physics of the cuprate superconductors, it is not a
comprehensive review, and it does not attempt to reflect the state
of the field by representing the variety of viewpoints that have
been taken elsewhere in the literature.

The primary assertions of this article are as follows. At the
lowest energy scales, the longest length scales, in the absence of
strong external perturbations, and at `optimal' carrier
concentrations and above, all experimental indications are that
the cuprate superconductors can indeed be described in the
framework of the BCS theory: the theory correctly captures the
primary order parameter of the superconducting state, and the
quantum numbers of its elementary excitations. However, many
experiments at lower doping concentrations and at shorter length
scales require one or more additional order parameters, either
conventional ({\em i.e.} associated with the breaking of a
symmetry of the Hamiltonian) or `topological' (see
Section~\ref{sec:topo} below). These order parameters are best
understood and classified in terms of the physics of ``Mott
insulators,'' a topic which will be discussed in greater detail
below. The importance of the Mott insulator was stressed by
\textcite{anderson87}. Our understanding of Mott insulators, and
of their classification into categories with distinct physical
properties has advanced greatly in the last decade, and a sharper
question of experimental relevance is: which class of Mott
insulators has its `order' present in the cuprate superconductors?
As we shall discuss below, the evidence so far supports a class
quite distinct from that implied in Anderson's proposal
\cite{sr91}.

\begin{figure}
\centerline{\includegraphics[width=3in]{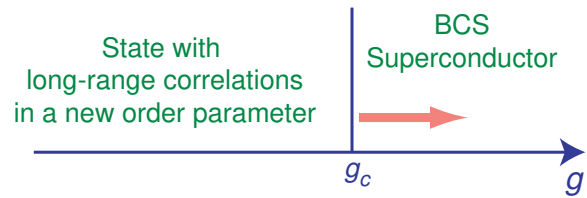}}
\caption{Our theoretical strategy for describing the influence of
a new order parameter in a BCS superconductor. Here $g$ is some
convenient coupling constant in the Hamiltonian, and we imagine
that the superconductor of physical interest is a BCS
superconductor with $g > g_c$. Theoretically, it is useful to
imagine that we can tune $g$ to a value smaller than $g_c$ where
there are long-range correlations in a new order parameter. Having
identified and understood the quantum phase transition at $g=g_c$,
we can expand away from it back towards the BCS superconductor (as
indicated by the thick arrow) to understand the influence of
quantum fluctuations of the new order parameter. This approach is
most effective when the transition at $g=g_c$ is second order, and
this will usually be assumed in our discussion. Note that the
horizontal axis need not be the concentration of mobile carriers,
and it may well be that the superconductor of physical interest
does not exhibit the $g<g_c$ state at any carrier concentration.}
\label{fig1}
\end{figure}

How can the postulated additional order parameters be detected
experimentally? In the simplest case, there could be long-range
correlations in the new order in the ground state: this is
apparently the case in La$_{2-\delta}$Sr$_{\delta}$CuO$_4$ at low
carrier concentrations, and we will describe recent experiments
which have studied the interplay between the new order and
superconductivity. However, the more common situation is that
there are no long range correlations in any additional order
parameter, but the `fluctuating' order is nevertheless important
in interpreting certain experiments. A powerful theoretical
approach for obtaining semi-quantitative predictions in this
regime of fluctuating order is provided by the theory of quantum
phase transitions: imagine that we are free to tune parameters so
that ultimately the new order does acquire long range correlations
somewhere in a theoretical phase diagram. A quantum critical point
will separate the phases with and without long-range order:
identify this critical point and expand away from it towards the
phase with fluctuating order, which is the regime of experimental
interest\cite{sy92,csy}; see Fig~\ref{fig1}.
An illuminating discussion of fluctuating order near quantum
critical points (along with a thorough analysis of many recent
experiments which has some overlap with our discussion here) has
been provided recently by \textcite{kivelson02}.

An especially important class of experiments involve perturbations
which destroy the superconducting order of the BCS state {\em
locally\/} (on the scale a few atomic spacings). Under such
situations the theory outlined above predicts that the order of
the Mott insulator is revealed in a halo surrounding the
perturbation, and can, in principle, be directly detected in
experiments. Perturbations of this type are Zn impurities
substituting on the Cu sites, and the vortices induced by an
applied magnetic field. We shall discuss their physics below.

To set the stage for confrontation between theory and experiment,
we review some essential features of the BCS theory in
Section~\ref{sec:bcs}, and introduce key concepts and order
parameters in the theory of Mott insulators in
Section~\ref{sec:mott}. We will combine these considerations in
our discussion of doped Mott insulators in Section~\ref{sec:dope},
which will also include a survey of some experiments. A
theoretical phase diagram which encapsulates much of the physics
discussed here appears in Section~\ref{sec:dope}, while
Section~\ref{sec:conc} concludes with a discussion on possible
directions for future work.

\section{BCS THEORY} \label{sec:bcs}

In BCS theory, superconductivity arises as an instability of a
metallic Fermi liquid. The latter state is an adiabatic
continuation of the free electron model of a metal, in which all
single particle states, labeled by the Bloch crystal momentum
$\vec{k}$, inside the $\vec{k}$-space Fermi surface are occupied
by electrons, while those outside remain empty. With
$c^{\dagger}_{\vec{k} \sigma}$ the creation operator for an
electron with momentum $\vec{k}$ and spin projection $\sigma =
\uparrow \downarrow$, a reasonable description of the Fermi liquid
is provided by the free electron Hamiltonian
\begin{equation}
H_0 = \sum_{\vec{k} \sigma} (\varepsilon_{\vec{k}} - \mu)
c^{\dagger}_{\vec{k} \sigma} c_{\vec{k} \sigma}, \label{h0}
\end{equation}
where $\varepsilon_{\vec{k}}$ is the energy-momentum dispersion of
the single-particle Bloch states and $\mu$ is the chemical
potential; the locus of points with $\varepsilon_{\vec{k}} = \mu$
defines the Fermi surface. Changes in electron occupation numbers
near the Fermi surface allow low energy processes which are
responsible for the conduction properties of metals.

BCS realized that an arbitrarily weak attractive interaction
between the electrons would induce the electrons near the Fermi
surface to lower their energy by binding into pairs (known as
Cooper pairs) \cite{cooper}. BCS also proposed a mechanism for
this attractive interaction: the exchange of a low energy phonon
between two electrons, along with the rapid screening of the
repulsive Coulomb interaction by the other electrons, leads to a
residual attractive interaction near the Fermi surface. We regard
this mechanism of electron pairing as a specific sidelight of BCS
theory for good metals, and not an essential characterization of
the BCS state. Indeed in liquid $^3$He, the pairing is believed to
arise from exchange of spin fluctuations (`paramagnons'), but the
resulting superfluid state has many key similarities to the
superconducting metals.

In the BCS ground state, the Cooper pairs undergo a process of
condensation which is very closely related to the Bose-Einstein
condensation of non-interacting bosons. Two well separated Cooper
pairs obey bosonic statistics when adiabatically exchanged with
each other, but their behavior is not simply that of point-like
Bose particles when their internal wavefunctions overlap---the
constituent electrons become important at these short scales;
however it is the long distance bosonic character which is crucial
to the appearance of a condensate of Cooper pairs. In the original
Bose-Einstein theory, the zero momentum boson creation operator
can be replaced by its $c$-number expectation value (due to the
occupation of this state by a macroscopic number of bosons);
similarly, the BCS state is characterized by the expectation value
of the creation operator of a Cooper pair with zero center of mass
momentum:
\begin{equation}
\left \langle c^{\dagger}_{\vec{k} \uparrow} c^{\dagger}_{-\vec{k}
\downarrow} - c^{\dagger}_{\vec{k} \downarrow}
c^{\dagger}_{-\vec{k} \uparrow} \right\rangle \propto
\Delta_{\vec{k}} \equiv \Delta_0 \left( \cos k_x - \cos k_y
\right). \label{pair}
\end{equation}
The functional form of (\ref{pair}) in spin and $\vec{k}$-space
carries information on the internal wavefunction of the two
electrons forming a Cooper pair: we have displayed a spin-singlet
pair with a $d$-wave orbital wavefunction on the square lattice,
as is believed to be the case in the cuprates
\cite{tsuei00,scalapino95}.

Along with (\ref{pair}) as the key characterization of the ground
state, BCS theory also predicts the elementary excitations. These
can be separated into two types: those associated with the motion
of center of mass, $\vec{R}$, of the Cooper pairs, and those in
which a pair is broken. The center of mass motion (or superflow)
of the Cooper pairs is associated with a slow variation in the
phase of the pairing condensate $\Delta_0 \rightarrow \Delta_0
e^{i \phi (\vec{R})}$: the superconducting ground state has $\phi
(\vec{R}) = $ a constant independent of $\vec{R}$ (and thus
long-range order in this phase variable), while a slow variation
leads to an excitation with superflow. A vortex excitation is one
in which this phase has a non-trivial winding, while the superflow
has a non-zero circulation:
\begin{equation}
\int_C d \vec{R} \cdot \nabla \phi = 2 \pi n_v \label{wind}
\end{equation}
where $n_v$ is the integer-valued vorticity, and $C$ is a contour
enclosing the vortex core. A standard gauge invariance argument
shows that each such vortex must carry a total magnetic flux of
$n_v hc /(2 e)$, where the $2e$ in the denominator represents the
quantum of charge carried by the ``bosons'' in the condensate.
Excitations which break Cooper pairs consist of multiple $S=1/2$
fermionic quasiparticles with dispersion
\begin{equation}
E_{\vec{k}} = \sqrt{(\epsilon_{\vec{k}} - \mu)^2 +
|\Delta_{\vec{k}}|^2}, \label{Ek}
\end{equation}
and these reduce to the particle and hole excitations around the
Fermi surface when $\Delta_0 \rightarrow 0$.

All indications from experiments so far are that the cuprate
superconductors do have a ground state characterized by
(\ref{pair}), and the elementary excitations listed above.
However, BCS theory does make numerous other predictions which
have been successfully and thoroughly tested in the low
temperature superconductors. In particular, an important
prediction is that if an external perturbation succeeds in
destroying superconductivity by sending $|\Delta_0| \rightarrow
0$, then the parent Fermi surface, which was swallowed up by the
Cooper instability, would reappear. This prediction is quite
different from the perspective discussed earlier, in which we
argued for the appearance of a halo of order linked to the Mott
insulator.

\section{MOTT INSULATORS} \label{sec:mott}

The Bloch theory of metals also specified conditions under which
crystalline materials can be insulating: if, after filling the
lowest energy bands with electrons, all bands are either fully
occupied or completely empty, then there is no Fermi surface, and
the system is an insulator. However, some materials are insulators
even though these conditions are not satisfied, and one-electron
theory would predict partially filled bands: these are Mott
insulators. Correlations in the motion of the electrons induced by
their Coulomb interactions are crucial in preventing metallic
conduction.

One of the parent compounds of the cuprate superconductors,
La$_2$CuO$_4$, is a simple example of a Mott insulator. The lowest
energy electronic excitations in this material reside on the Cu
$3d_{x^2-y^2}$ orbitals, which are located on the vertices of a
square lattice. The crystal has a layered structure of stacked
square lattices, with only a weak amplitude for electron hopping
between successive layers. (We shall neglect the interlayer
coupling and focus on the physics of a single square lattice in
the remainder of this article.) After accounting for the
ionization states of the other ions in La$_2$CuO$_4$, there turns
out to be exactly one electron per unit cell available to occupy
the Cu $3d_{x^2-y^2}$ band. With two available spin states, this
band can accommodate two electrons per unit cell, and so is
half-filled, and should have a metallic Fermi surface.
Nevertheless, La$_2$CuO$_4$ is a very good insulator. The reason
for this insulating behavior can be understood quite easily from a
simple classical picture of electron motion in the presence of the
Coulomb interactions. Classically, the ground state consists of
one electron localized on each of the $3d_{x^2-y^2}$ orbitals:
this state minimizes the repulsive Coulomb interaction energy. Any
other state would have at least one orbital with two electrons,
and one with no electrons: there is a large energetic penalty for
placing two electrons so close to each other, and this prohibits
motion of electrons across the lattice: hence the Mott insulator.

\begin{figure}
\centerline{\includegraphics[width=1.8in]{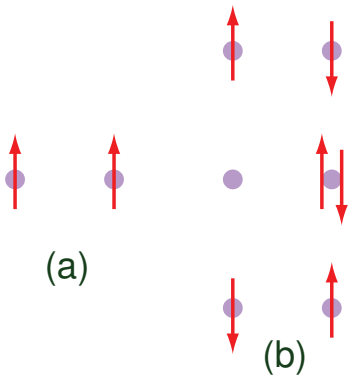}}
\caption{Motion of the two ferromagnetically aligned spins in
({\em a\/}) is prohibited by the Pauli principle. In contrast, the
antiferromagnetically aligned spins at the top and bottom in ({\em
b\/}) can access a high energy intermediate state (shown in the
middle of ({\em b\/})) and so undergo an exchange process.}
\label{fig2}
\end{figure}

Let us now look at the quantum theory of the Mott insulator more
carefully. While charge fluctuations on each site are expensive,
it appears that the spin of the electron can be rotated freely and
independently on each site. However, in the quantum theory virtual
charge fluctuations do occur, and these lead to residual
``super-exchange'' interactions between the spins
\cite{anderson59}. We represent the spin on the Cu site $j$ by the
$S=1/2$ spin operator ${\bf S}_j$; the effective Hamiltonian that
describes the spin dynamics then takes the form
\begin{equation}
H = \sum_{i<j} J_{ij} {\bf S}_i \cdot {\bf S}_j + \ldots
\label{afm}
\end{equation}
where the $J_{ij}$ are short-ranged exchange couplings and the
ellipses represent possible multiple spin couplings, all of which
preserve full SU(2) spin rotation invariance. Because the Pauli
principle completely prohibits charge fluctuations between two
sites if they have parallel spin electrons, while they are only
suppressed by the Coulomb repulsion if they have opposite spins
(see Fig~\ref{fig2}), we expect an antiferromagnetic sign $J_{ij}
> 0$, so that nearby spins prefer opposite orientations.
Classifying quantum ground states of models like (\ref{afm}) is a
problem of considerable complexity, and has been the focus of
extensive research in the last decade. We summarize the current
understanding below.

In keeping with the spirit of this article, we characterize ground
states of $H$ by a number of distinct order parameters. We only
discuss states below which have long-range correlation in a single
order parameter; in most cases, co-existence of multiple order
parameters is also allowed \cite{balents99,sf00}, but we will
ignore this complexity here. Our list of order parameters is not
exhaustive, and we restrict our attention to the most plausible
candidates (in the author's opinion) for short-range
$J_{ij}$.\footnote{An order that has been much discussed in the
literature, which we do not discuss here, is that associated with
the staggered flux state \protect\cite{affleck88}, and the related
algebraic spin liquid \protect\cite{rantner01,wen02}. The low
energy theory of these states includes a gapless U(1) gauge field,
and it has been argued \protect\cite{sp02} that instantons, which
are allowed because the underlying lattice scale theory has a
compact gauge symmetry, always proliferate and render these
unstable towards confining states (of the type discussed in
Section~\protect\ref{sec:bond}) in two spatial dimensions.
However, states with a gapless U(1) gauge field are allowed in
three spatial dimensions \cite{motrunich02,wen02a}.}

Although our discussion below will refer mainly to Mott
insulators, we will also mention ground states of non-insulating
systems with mobile charge carriers: the order parameters we use
to characterize Mott insulators can be applied more generally to
other systems, and this will done in more detail in
Section~\ref{sec:dope}.

\subsection{Magnetically ordered states}
\label{sec:mag} Such states are obtained by examining $H$ for the
case of large spin $S$ on each site: in this limit, the ${\bf
S}_j$ can be taken as classical $c$-numbers, and these take a
definite non-zero value in the ground state. More precisely, the
SU(2) spin rotation symmetry of $H$ is spontaneously  broken in
the ground state by the non-zero values of $\langle {\bf S}_j
\rangle$, which are chosen to minimize the energy of $H$. We
consider only states without a net ferromagnetic moment ($\sum_j
\langle {\bf S}_j \rangle = 0$), and this is expected because
$J_{ij} > 0$. The pattern of non-zero $\langle {\bf S}_j \rangle$
can survive down to $S=1/2$, and this is often found to be the
case, although quantum fluctuations do significantly reduce the
magnitude of $\langle {\bf S}_j \rangle$.

An especially important class of magnetically ordered
states\footnote{Magnetically ordered states with the values of
$\langle {\bf S}_j \rangle$ non-coplanar ({\em i.e.\/} three
dimensional spin textures) are not included in this simple
classification. Their physical properties are expected to be
similar to those of the non-collinear case discussed in
Section~\protect\ref{sec:noncoll} in that quantum fluctuations of
such a state lead to a paramagnet with topological order. However,
this paramagnet is likely to have also a broken time-reversal
symmetry.} is characterized by a single ordering wavevector
$\vec{K}$:
\begin{equation}
\langle {\bf S}_j \rangle = {\bf N}_1 \cos \left(\vec{K} \cdot
\vec{r}_j \right) + {\bf N}_2 \sin \left(\vec{K} \cdot \vec{r}_j
\right) \label{spin}
\end{equation}
where $\vec{r}_j$ is the spatial location of the site $j$, and
${\bf N}_{1,2}$ are two fixed vectors in spin space. We list two
key subcategories of magnetically ordered Mott insulators which
obey (\ref{spin}):

\subsubsection{Collinear spins, ${\bf N}_1 \times {\bf N}_2 = 0$}
\label{sec:collinear} 
\begin{figure}
\centerline{\includegraphics[width=3in]{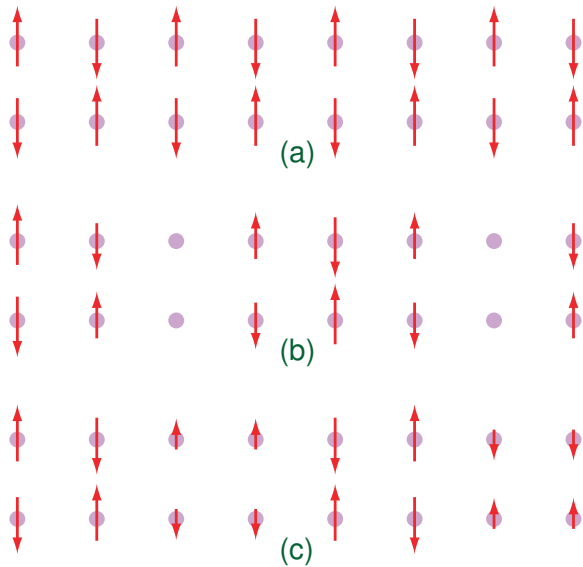}}
\caption{States with collinear magnetic order on a square lattice
with unit lattice spacing and wavevectors ({\em a\/}) $\vec{K} =
(\pi,\pi)$, ({\em b\/}) and ({\em c\/}) $\vec{K} = (3\pi/4,\pi)$.
Shown are the values of (\protect\ref{spin}) on the square lattice
sites ${\bf r}_j$. A single unit cell is shown for the latter two
states; they are crystographically inequivalent and have different
reflection planes: in ({\em b\/}) the reflection planes are on
certain sites, while in ({\em c\/}) they are at the midpoint
between two sites.} \label{fig3}
\end{figure}

In this situation, the mean values of the
spins in (\ref{spin}) on all sites $j$ are either parallel or
anti-parallel to each other. The undoped insulator La$_2$CuO$_4$
is of this type\footnote{For this special value of $\vec{K}$ on
the square lattice, and with the origin of ${\bf r}$ co-ordinates
on a lattice site, (\protect\ref{spin}) is actually independent of
${\bf N}_2$)} with $\vec{K} = (\pi, \pi)$; see Fig~\ref{fig3}a.
Insulating states with static holes appeared in
\textcite{zaanen89}, \textcite{machida89}, \textcite{schulz89},
and \textcite{poilblanc89} with ordering wavevectors which move
continuously away from $(\pi,\pi)$.
Another important illustrative example is the case $\vec{K} = (3
\pi/4, \pi)$. Such a wavevector could be preferred in a Mott
insulator by longer range $J_{ij}$ in (\ref{afm}), but in practice
it is found in a non-insulating state obtained by doping
La$_2$CuO$_4$ with a suitable density of mobile carriers
\cite{tranquada95,white98,white99,white98a,wakimoto99,wakimoto01,kivelson96,seibold98,martin00}---we
can crudely view the mobile carriers as having induced an
effective longer range exchange between the spins. Two examples of
states with this value of $\vec{K}$ are shown in Fig~\ref{fig3}, a
{\em site}-centered state with ${\bf N}_2 = 0$ in Fig~\ref{fig3}b,
and a {\em bond}-centered state with ${\bf N}_2 = (\sqrt{2}-1){\bf
N}_1$ in Fig~\ref{fig3}c. (The states have planes of reflection
symmetry located on sites and the centers of bonds respectively,
and so are crystallographically inequivalent. Also, these
inequivalent classes are only present if the wavevector ${\bf K}$
is commensurate with the underlying lattice.)

\subsubsection{Non-collinear spins, ${\bf N}_1 \times {\bf N}_2
\neq 0$}
\label{sec:noncoll}
\begin{figure}
\centerline{\includegraphics[width=3in]{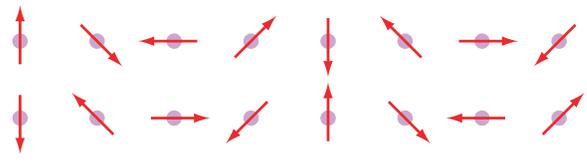}} \caption{A
state with non-collinear magnetic order on the square lattice
defined by (\ref{spin}) and (\ref{spiral}) with wavevector
$\vec{K} = (3\pi/4,\pi)$.} \label{fig4}
\end{figure}

Now the spin expectation values in
(\ref{spin}) lie in a plane in spin space, rather than along a
single direction. For simplicity, we will only consider the
simplest, and most common, case of non-collinearly ordered state,
in which
\begin{equation}
{\bf N}_1 \cdot {\bf N}_2 = 0~~~;~~~{\bf N}_1^2 = {\bf N_2}^2 \neq
0, \label{spiral} \
\end{equation}
and then the values of $\langle {\bf S}_j \rangle$ map out a
circular spiral \cite{shraiman88,shraiman89}, as illustrated in
Fig~\ref{fig4}.

\subsection{Paramagnetic states}
\label{sec:para} The other major class of states comprises those
having
\begin{equation}
\langle {\bf S}_j \rangle = 0,
\end{equation}
and the ground state is a total spin singlet.\footnote{In a finite
system with an even number of spins, the magnetically ordered
ground state also has total spin zero. However, to obtain a state
which breaks spin rotation symmetry as in (\protect\ref{spin}), it
is necessary to mix in a large number of nearly degenerate states
which carry non-zero total spin. The paramagnetic does not have
such higher spin states available at low energy in a finite
system.} Loosely speaking each spin ${\bf S}_j$ finds a partner,
say ${\bf S}_{j'}$, and the two pair up to form a singlet valence
bond
\begin{equation}
\frac{1}{\sqrt{2}}\left( |\uparrow \rangle_j |\downarrow
\rangle_{j'} - |\downarrow \rangle_j |\uparrow \rangle_{j'}
\right). \label{singlet}
\end{equation}
Of course, there are many other choices for the partner of spin
${\bf S}_j$, and in the Feynman path integral picture we imagine
that the pairing configuration fluctuates in quantum imaginary
time; this is the `resonating valence bond' picture of
\textcite{pauling49}, \textcite{fazekas74}, and
\textcite{anderson87}. However, there is a great of structure and
information contained in the manner in which this fluctuation
takes place, and research \cite{rs91,sr91,css94,css94a}
dilineating this structure has led to the following classification
of paramagnetic Mott insulators.

\subsubsection{Bond-ordered states: confined spinons}
\label{sec:bond} 
\begin{figure}
\centerline{\includegraphics[width=3in]{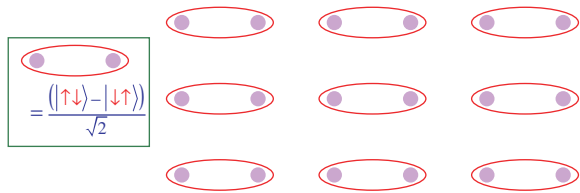}} \caption{A
crude variational wavefunction of a bond-ordered paramagnetic
state. The true ground state will have fluctuations of the singlet
bonds about the configuration shown here, but its pattern of
lattice symmetry breaking will be retained. In other words, each
bond represented by an ellipse above will have the same value of
$\langle Q_{a} (\vec{r}_j) \rangle$, and this value will be
distinct from that associated with all other bonds. This pattern
of symmetry breaking is represented more abstractly in
Fig~\protect\ref{fig6}a. } \label{fig5}
\end{figure}

This class of states can be easily understood by
the caricature of its wavefunction shown in Fig~\ref{fig5}: here
each spin has chosen its valence bond partner in a regular manner,
so that there is a long-range `crystalline' order in the
arrangement of valence bonds.
This ordering of bonds clearly breaks the square lattice space
group symmetries under which the Hamiltonian is invariant. Of
course, the actual wavefunction for any realistic Hamiltonian will
have fluctuations in its valence bond configuration, but the
pattern of lattice symmetry breaking implied by Fig~\ref{fig5}
will be retained in the true bond-ordered ground state. We can
make this precise by examining observables which are insensitive
to the electron spin direction: the simplest such observables we
can construct from the low energy degrees of freedom of the Mott
insulator are {\em bond} variables, which are a measure of the
exchange energy between two spins:
\begin{equation}
Q_a (\vec{r}_j ) \equiv {\bf S}_j \cdot {\bf S}_{j+a}.
\label{bond}
\end{equation}
Here $a$ denotes displacement by the spatial vector $\vec{r}_a$,
and so the spins above are at the spatial locations $\vec{r}_j$
and $\vec{r}_j + \vec{r}_a$. We will mainly consider bond order
with $\vec{r}_a \neq 0$, but note that the on-site variable $Q_0
(\vec{r}_j) $, with $\vec{r}_a = 0$, is a measure of the charge
density\footnote{By (\protect\ref{bond}), $Q_0 (\vec{r}_j) = {\bf
S}_j^2$. A site with a spin has ${\bf S}_j^2 = 3/4$, while a site
with a hole has ${\bf S}_j^2 = 0$, and we assume that doubly
occupied sites are very rare. Thus ${\bf S}_j^2$, and hence $Q_0
(\vec{r}_j)$ is seen to be linearly related to the charge density
on site $j$.} on site $\vec{r}_j$, and so this special case of
(\ref{bond}) measures the ``charge order.''

\begin{figure}
\centerline{\includegraphics[width=3in]{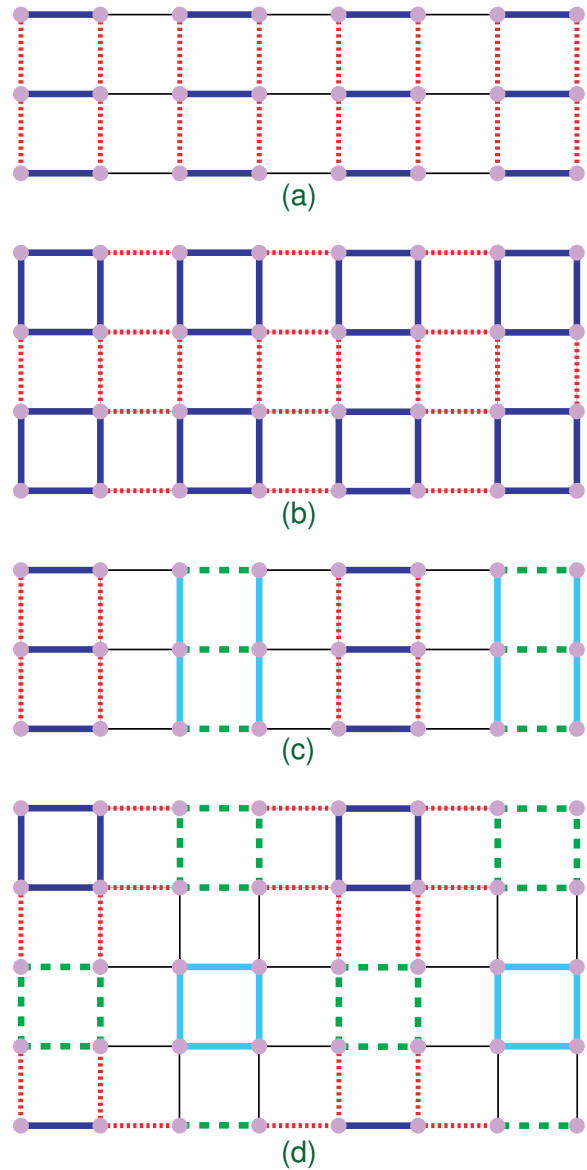}}
\caption{Pattern of the bond variables $\langle Q_a (\vec{r}_j )
\rangle$, for $\vec{r}_a$ a nearest-neighbor vector, in a number
of paramagnetic states with $\langle {\bf S}_j \rangle = 0$. For
each state, the values of $\langle Q_a (\vec{r}_j ) \rangle$ are
equal on bonds represented by the same type of line, and unequal
otherwise. The number of distinct values of $\langle Q_a
(\vec{r}_j ) \rangle$ are ({\em a \/}) 3, ({\em b \/}) 2, ({\em c
\/}) 5, and ({\em d \/}) 5. The unit cells of the ground states
have sizes ({\em a \/}) $2 \times 1$, ({\em b \/}) $2 \times 2$,
({\em c \/}) $4 \times 1$, and ({\em d \/}) $4 \times 4$.}
\label{fig6}
\end{figure}

The state introduced in Fig~\ref{fig5} can be characterized by the
pattern of values of $\langle Q_a (\vec{r}_j ) \rangle$ with
$\vec{r}_a$ a nearest neighbor vector, as shown in
Fig~\ref{fig6}a: notice there are 3 distinct values of $\langle
Q_a (\vec{r}_j ) \rangle$ and symmetries of the states in
Figs~\ref{fig5} and~\ref{fig6}a are identical.
While these 3 values are quite different in the trial state in
Fig~\ref{fig5}, their values in the actual ground state may be
quite close to each other: it is only required that they not be
exactly equal.

Another closely related bond-ordered state, which has appeared in
some theories \cite{rs89a,dombre89,sr96,altman02}, is shown in
Fig~\ref{fig6}b: here the bonds have a plaquette-like arrangement
rather than columnar, but, as we shall discuss below, the physical
properties of all the states in Fig~\ref{fig6} are quite similar
to each other.

We can also consider patterns of bond order with larger unit
cells, and two important structures which have appeared in
theories of doped Mott insulators \cite{vs99,vojta02} are shown in
Figs~\ref{fig6}c and~\ref{fig6}d (related bond orders also appear
in studies of quasi-one-dimensional models appropriate to organic
superconductors \cite{mazumdar00,clay02}). Again, as in
Section~\ref{sec:collinear}, such states could, in principle, also
appear in Mott insulators with longer-range exchange in
(\ref{afm}). An interesting property of these states is that,
unlike the states in Fig~\ref{fig6}a and~\ref{fig6}b, not all
sites are crystallographically equivalent. This means that on-site
spin-singlet observables, such as the site charge density, will
also have a spatial modulation from site to site. A subtlety is
that the Hamiltonian (\ref{afm}) acts on a Hilbert space of
$S=1/2$ spins on every site, and so the charge density on each
site is fixed at unity. However, it must be remembered that
(\ref{afm}) is an effective model derived from an underlying
Hamiltonian which does allow virtual charge fluctuations, and the
site charge modulations in the states of Figs~\ref{fig6}c
and~\ref{fig6}d will appear when it is properly computed in terms
of the microscopic degrees of freedom. At the same time, this
argument also makes it clear that any such modulation is
suppressed by the repulsive Coulomb energy, and could well be
difficult to observe, even in the doped antiferromagnet. So the
on-site variable, $\langle Q_{0} (\vec{r}_j) \rangle = \langle
{\bf S}_j^2 \rangle$, will have a weak modulation in the states of
Fig~\ref{fig6}c and~\ref{fig6}d when computed in the full Hilbert
space of the model with charge fluctuations. Note, however, that
the modulation in bond orders associated with $Q_a (\vec{r}_j )$,
with $\vec{r}_a \neq 0$, need not be small in the states in
Fig~\ref{fig6}, as such modulations are not suppressed as
effectively by the Coulomb interactions.

\begin{figure}
\centerline{\includegraphics[width=3in]{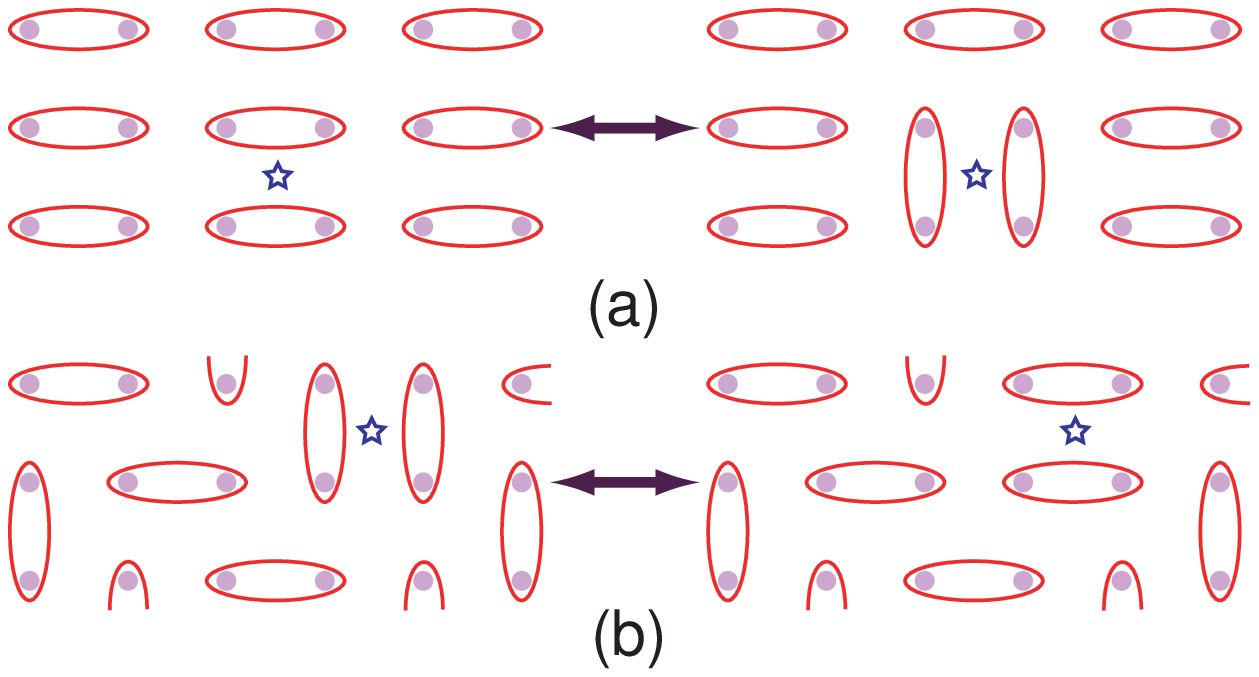}}
\caption{Bond order induced by quantum fluctuations. Valence bonds
gain energy by ``resonating'' in pairs
\protect\cite{pauling49,fazekas74,anderson87,rk88}; shown are
resonances around the plaquette ({\em i.e.\/} square loop) marked
with a star. For the regular bond-ordered configuration of valence
bonds in (a), such resonance can occur not only around the
plaquette marked with a star, but around five additional
plaquettes. In contrast, in (b), such a resonance is possible only
around the plaquette marked with a star. This additional quantum
``entropy'' associated with (a) selects regular bond order in the
ground state. More sophisticated considerations (which also allow
valence bonds that do not connect nearest neighbor sites) show
that this mechanism is especially effective in two dimensions
\protect\cite{rs90,sp02}.} \label{fig7}
\end{figure}

The physical mechanism inducing bond-ordered states such as those
in Fig~\ref{fig6} is illustrated in the cartoon pictures in
Fig~\ref{fig7}.
More detailed computations rely on a semiclassical theory of
quantum fluctuations near a magnetically ordered state
\cite{rs90}. Remarkably, very closely related theories also appear
from a very different starting point---from duality mappings
\cite{rs90,fradkin90} of ``quantum dimer models'' \cite{rk88} of
the paramagnetic state. These computations show that spontaneous
bond order invariably appears in the ground state in systems with
collinear spin correlations in two spatial
dimensions\cite{rs90,sp02}. We will have more to say about this
connection between and bond and collinear spin order in
Section~\ref{sec:collbond}.

We also mention here the ``nematic'' states of \textcite{kfe98} in
the doped Mott insulator. These can also be characterized by the
bond order variables in (\ref{bond}). The symmetry of translations
with respect to $\vec{r}_j$ is not broken in such states, but the
values of $\langle Q_{a} (\vec{r}_j) \rangle$ for symmetry-related
values of $\vec{r}_a$ become unequal. For example, $\langle Q_{a}
(\vec{r}_j )\rangle$ has distinct values for $\vec{r}_a = (1,0)$
and $(0,1)$. Such states also appear in certain insulating
antiferromagnets \cite{rs89a,rs89b,rs90}.

It also interesting to note here that the bond order variables
$Q_a (\vec{r}_j )$ also have spatial modulations in some of the
magnetically ordered states considered in Section~\ref{sec:mag}
\cite{zachar98}. It is clear from (\ref{bond}) that any broken
lattice symmetry in the spin-rotation invariant quantity $\langle
{\bf S}_j \rangle \cdot \langle {\bf S}_{j+a} \rangle$ will
generate a corresponding broken symmetry in the bond variable
$\langle Q_a (\vec{r}_j ) \rangle$. Evaluating the former using
(\ref{spin}) we can deduce the following: ({\em i}) the
$\vec{K}=(\pi,\pi)$ state in Fig~\ref{fig3}a and the spiral state
in Fig~\ref{fig4} have $\langle Q_a (\vec{r}_j ) \rangle$
independent of $\vec{r}_j$, and hence no bond order; ({\em ii})
the bond-centered magnetically ordered state in Fig~\ref{fig3}c
has precisely the same pattern of bond order as the paramagnetic
state in Fig~\ref{fig6}c; ({\em iii}) the site-centered
magnetically ordered state in Fig~\ref{fig3}b has bond order with
$\langle Q_a (\vec{r}_j ) \rangle$ $\vec{r}_j$-dependent, but with
a pattern distinct from any shown here---this pattern of bond
order is in principle also allowed for paramagnetic states, but
has so far not been found to be stable in various studies.
Finally, note that in ({\em ii}) and ({\em iii}) the period of the
bond order (four) is half that of the spin modulation
(eight)--this is easily seen to be a general relationship
following from the correspondence $\langle Q_a (\vec{r}_j )
\rangle \sim \langle {\bf S}_{j} \rangle \cdot \langle {\bf
S}_{j+a} \rangle + \ldots$ in magnetically ordered states, which
with (\ref{spin}) implies an $\vec{r}_j$-dependent modulation of
the bond order with wavevector $2 \vec{K}$. It is worth
reiterating here that this last relationship should not be taken
to imply that there are no modulations in $\langle Q_a (\vec{r}_j
) \rangle$ when $\langle {\bf S}_a \rangle = 0$: there can indeed
be bond modulations in a paramagnet, as discussed in the other
paragraphs of this subsection, and as is already clear from the
simple wavefunction in Fig~\ref{fig5}---these will be important
later for physical applications.

\begin{figure}
\centerline{\includegraphics[width=3in]{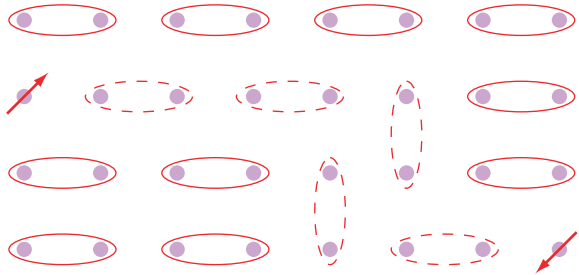}}
\caption{Linear confining potential between two neutral $S=1/2$
spinons in a bond-ordered state. The line of valence bonds with
dashed lines is out of alignment with the global bond order, and
it costs a finite energy per unit length.} \label{fig8}
\end{figure}

We continue our exposition of paramagnetic bond-ordered states by
describing excitations with non-zero spin. These can be understood
simply by the analog of cartoon wavefunction pictures drawn in
Fig~\ref{fig5}. To create free spins we have to break at least one
valence bond, and this initially creates two unpaired, neutral,
$S=1/2$ degrees of freedom (the ``spinons''). We can ask if the
spinons can be moved away from each other out to infinity, thus
creating two neutral $S=1/2$ quasiparticle excitations. As
illustrated in Fig~\ref{fig8}, this is not the case: connecting
the two spinons is a line of defect valence bonds which is not
properly aligned with the global bond order, and these defects
have a finite energy cost per unit length.
This linearly increasing potential is quite analogous to that
between a quark and an anti-quark in a meson, and the spinons
(quarks) are therefore permanently confined \cite{rs89b}. Moving
two spinons apart from each other will eventually force the
breaking of the defect line by the creation of another pair of
spinons. The only stable excitation with nonzero spin therefore
consists of a pair of spinons and carries spin $S=1$. We will
refer to this quasiparticle as a {\em spin exciton} as its quantum
numbers and observable characteristics are quite similar to spin
excitons found in semiconductors and metals. The spin exciton is
clearly the analog of a meson consisting of a quark and anti-quark
pair.

\begin{figure}
\centerline{\includegraphics[width=3in]{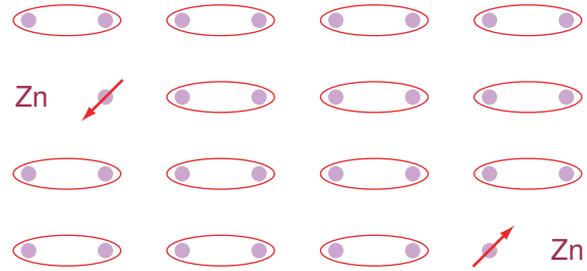}}
\caption{Cartoon wavefunction for 2 static spinless Zn impurities
in a confining, bond-ordered state. It we attempt to construct a
wavefunction only using singlet valence bonds, then just as in
Fig~\protect\ref{fig8}, there will be defect line of singlet bonds
which are not aligned with the global bond order, which will cost
a finite energy per unit length. When the two Zn impurities are
sufficiently far apart, it will pay to restore the bond order in
between the impurities, at the price of unpaired $S=1/2$ moments,
one near each impurity.} \label{fig9}
\end{figure}

A similar reasoning can be used to understand the influence of
static spinless impurities {\em i.e.} the consequences of removing
a $S=1/2$ spin from a fixed site $j$ in (\ref{afm}).
Experimentally, this can be conveniently done by substituting a
spinless Zn$^{++}$ ion in place of an $S=1/2$ Cu ion. The main
physical effect can be understood from the cartoon wavefunction in
Fig~\ref{fig9}: it is convenient to imagine placing 2 Zn
impurities, and then moving them apart out to infinity to deduce
the physics in the vicinity of a single impurity.
As in our discussion above for spinons, note that there will
initially be a line of defect valence bonds connecting the two Zn
impurities, but it will eventually pay to annihilate this defect
line by creating two spinons and binding each to a Zn impurity.
Thus each Zn impurity {\em confines} a free $S=1/2$ spinon in its
vicinity, and this can be detected in experiments
\cite{fink90}.\footnote{In principle the Zn impurity could also
bind an electron (with or without a spinon) but this is suppressed
by the charge gap in a Mott insulator. Later, in
Section~\protect\ref{sec:imp} when we consider Zn impurities in
$d$-wave superconductors, a related phenomenon appears in the form
of the Kondo effect.}

\subsubsection{Topological order: free spinons}
\label{sec:topo} 
\begin{figure}
\centerline{\includegraphics[width=3in]{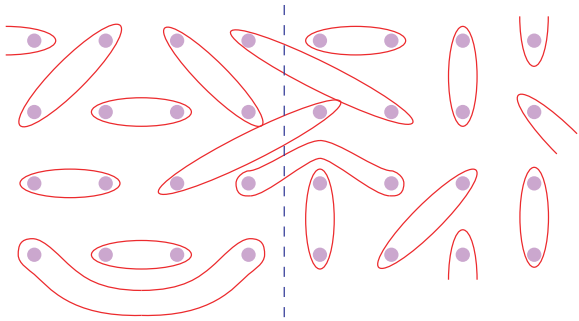}}
\caption{Topological order in a resonating valence bond state.
Shown is one component of the wavefunction, with a particular
pairing of the spins into local singlets: the actual wavefunction
is a superposition over a very large number of such pairing
configurations. The number of valence bonds cutting the dashed
line is an invariant modulo 2 over these pairing configurations,
as shown by the following simple argument. Any rearrangement of
the valence bonds can be reached by repeated application of an
elementary rearrangement between 4 spins: $(1,2)(3,4) \rightarrow
(1,3)(2,4)$ (here (i,j) denotes a singlet bond between ${\bf S}_i$
and ${\bf S}_j$). So it is sufficient to check this conservation
law for 4 spins: this is done easily by explicitly considering all
different possibilities among spins 1,2,3,4 residing to the
left/right of the dashed line. If the system has periodic boundary
conditions along the horizontal direction, then this conservation
law is violated, but only by rearrangements associated with loops
which circumnavigate the systems; these only occur with a
probability which becomes exponentially small as the circumference
of the system increases.} \label{fig10}
\end{figure}

This type of paramagnet is the ``resonating
valence bond'' (RVB) state
\cite{pauling49,fazekas74,anderson87,baskaran88,moessner01,krs87}
in which the singlet pairings fluctuate in a liquid-like
configuration,\footnote{In recent years,
\protect\textcite{anderson02} has extended the RVB concept to
apply to doped Mott insulators at temperatures {\em above} $T_c$.
This extension is not in consonance with the classification of the
present article. The topological order discussed in this
subsection can only be defined at $T=0$ in two spatial dimensions.
The description at $T>T_c$ requires solution of a problem of
quantitative difficultly, and with incoherent excitations, but
without sharp distinctions between different states.} in contrast
to the crystalline arrangement in Fig~\ref{fig5}. Despite the
apparent `disorder' in the valence bond configuration in the
ground state, there is actually a subtle topological order
parameter which characterizes this type of Mott insulator
\cite{thouless87,bonesteel89,rk88,kivelson89,rc89,wen91,rs91}, and
which plays an important role in determining its excitation
spectrum. The reader can see this in the context of the cartoon
picture shown in Fig~\ref{fig10}.
Count the number of singlet valence bonds cutting the dashed line
in this figure: this number will clearly depend upon the
particular valence bond configuration chosen from the many present
in the ground state, and one such is shown in Fig~\ref{fig10}.
However, as argued in the figure caption, the number of bonds
cutting the dashed line is conserved modulo 2 between any two
configurations which differ only local rearrangements of valence
bonds: the quantum number associated with this conservation is the
topological order in the ground state.

A convenient and powerful description of this topological order is
provided by an effective model of the singlet sector formulated as
$Z_2$ gauge theory \cite{rs91,sr91,wen91,sf00}.\footnote{Readers
not familiar with $Z_2$ gauge theories may understand them by
analogy to electromagnetism. The latter is a U(1) gauge theory in
which the physics is invariant under the transformation $z
\rightarrow e^{i\phi}z$, $A_{\mu} \rightarrow A_{\mu} -
\partial_{\mu} \phi$ where $z$ is some matter field, $A_{\mu}$ is
a gauge field, and $\phi$ is an arbitrary spacetime-dependent
field which generates the gauge transformation. Similarly, in a
$Z_2$ gauge theory, matter fields transform as $z \rightarrow \eta
z$, where $\eta$ is a spacetime-dependent field which generates
the gauge transformation, but is now allowed to take only the
values $\eta = \pm 1$. The $Z_2$ gauge field $\sigma_{ij}$ resides
on the links of a lattice, and transforms as $\sigma_{ij}
\rightarrow \eta_i \sigma_{ij} \eta_j$.} We postpone a
self-contained derivation of this $Z_2$ gauge theory to
Section~\ref{sec:noncolltopo} (see especially Fig~\ref{fig11}):
here, we show that such a gauge theory has similar topological
properties. In a system with periodic boundary conditions (with
the topology of a torus), the $Z_2$ gauge theory has different
sectors depending upon whether there is a $Z_2$ flux piercing any
of the holes of the torus (following \textcite{sf00}, this $Z_2$
flux is now commonly referred to as a ``vison''). In the valence
bond picture discussed in the previous paragraph, a vison changes
the sign associated with every valence bond cutting a line
traversing the system in the vison direction (the dashed line in
Fig~\ref{fig9}); in other words, the even and odd valence bond
sectors mentioned above now have their relative signs in the
wavefunction changed.

In addition to appearing in the holes of the torus, the vison can
also appear as a singlet excitation within the bulk
\cite{kivelson89,rc89,rs91,sf00}. It is now a vortex excitation in
the $Z_2$ gauge theory, that requires a finite energy for its
creation. We will see below in Section~\ref{sec:noncolltopo} that
there is an alternative, and physically revealing, interpretation
of this vortex excitation in terms of the order parameters used
earlier to characterize the magnetically ordered state, and that
the topological order is intimately connected to the vison energy
gap.

Finally, we can describe the spin-carrying excitations of this
topologically ordered state using the crude, but instructive,
methods used in Section~\ref{sec:bond}. As there is no particular
bond order associated with the ground state, the spinons have no
confining force between them, and are perfectly free to travel
throughout the system as independent neutral $S=1/2$
quasiparticles. Similarly, there is no confining force between Zn
impurities and the spinons, and so it is not required that an
$S=1/2$ moment be present near each Zn impurity
\cite{icmp,fendley02}.

\subsection{Connections between magnetically ordered and
paramagnetic states} \label{sec:connect}

A central ingredient in the reasoning of this article is the claim
that there is an intimate connection between the magnetically
ordered states in Section~\ref{sec:mag} and a corresponding
paramagnetic state in Section~\ref{sec:para}. In particular, the
collinear states of Section~\ref{sec:collinear} are linked to the
bond-ordered states in Section~\ref{sec:bond}, while the
non-collinear states of Section~\ref{sec:noncoll} are linked to
the topologically ordered states of Section~\ref{sec:topo}. The
reader will find a more technical discussion of the following
issues in a companion review article by the author
\cite{ssannals}.

Before describing these links in the following subsections, we
discuss the meaning of the ``connectedness'' of two states. The
magnetically ordered phases are characterized by simple order
parameters that we have discussed in Section~\ref{sec:mag}. Now
imagine a second-order quantum phase transition in which the
magnetic long-range order is lost, and we reach a state with
fluctuating magnetic correlations, which is ultimately a
rotationally invariant, spin-singlet paramagnet at the longest
length scales. We will review arguments below which show that this
``quantum disordered'' state \cite{chn} is characterized by the
order parameter of the connected paramagnetic state {\em i.e.}
fluctuating collinear magnetic order leads to bond order, while
fluctuating non-collinear magnetic order can lead to topological
order. So two connected states are generically proximate to each
other, without an intervening first order transition, in a
generalized phase diagram drawn as a function of the couplings
present in the Hamiltonian.

\subsubsection{Collinear spins and bond order}
\label{sec:collbond}

It should be clear from Section~\ref{sec:collinear} that collinear
spin states are characterized by a single vector ${\bf N}_1$. The
second vector ${\bf N}_2$ is pinned to a value parallel to ${\bf
N}_1$ by some short distance physics, and at long distances we may
consider a theory of the fluctuations of ${\bf N}_1$ alone. In a
phase with magnetic order, the dominant spin-wave fluctuations
occur in configurations with a fixed non-zero value of $|{\bf N}_1
|$. In the transition to a non-magnetic phase, the mean value of
$|{\bf N}_1 |$ will decrease, until the fluctuations of ${\bf
N}_1$ occur about ${\bf N}_1 = 0$ in a paramagnetic phase. There
are 3 normal modes in this fluctuation spectrum, corresponding to
the 3 directions in spin space, and the resultant is an $S=1$
gapped quasiparticle excitation in the paramagnetic state. This we
can easily identify as the $S=1$ spin exciton of the bond-ordered
state: this identification is evidence supporting our claimed
connection between the states of Section~\ref{sec:collinear} and
~\ref{sec:bond}.

Further evidence is provided by detailed computations which show
the appearance of bond order in the regime where ${\bf N}_1$
fluctuations have lost their long-range order. We have already
seen a simple example of this above in that the magnetically
ordered state in Fig~\ref{fig3}c already had the bond order of the
paramagnetic state in Fig~\ref{fig6}c: it is completely natural
for the bond order in the magnetically ordered phase in
Fig~\ref{fig3}c to persist across a transition in which spin
rotation invariance is restored, and this connects it to the state
in Fig~\ref{fig6}c. A non-trivial example of a related connection
is that between the $\vec{K} = (\pi, \pi)$ N\'{e}el state in
Fig~\ref{fig3}a, and the paramagnetic bond-ordered states in
Figs~\ref{fig6}a and~\ref{fig6}b, which was established by
\textcite{rs89b}, \textcite{rs90}, and \textcite{sp02}: Berry
phases associated with the precession of the lattice spins were
shown, after a duality mapping, to induce bond order in the phase
in which long-range order in ${\bf N}_1$ was lost.

\subsubsection{Non-collinear spins and topological order}
\label{sec:noncolltopo}

The first argument of Section~\ref{sec:collbond}, when generalized
to non-collinear spins, leads quite simply to a surprisingly
subtle characterization of the associated paramagnetic phase.

Recall from Section~\ref{sec:noncoll} that the non-collinear
magnetic phase is characterized by two orthogonal, and equal
length, vectors ${\bf N}_{1,2}$. It takes 6 real numbers to
specify two vectors, but the 2 constraints in (\ref{spiral})
reduce the number of real parameters required to specify the
ordered state to 4. There is a useful parameterization
\cite{css94,css94a} which explicitly solves the constraints
(\ref{spiral}) by expressing ${\bf N}_{1,2}$ in terms of 2 complex
numbers $z_{\uparrow}$, $z_{\downarrow}$ (which are equivalent to
the required 4 real numbers):
\begin{equation}
{\bf N}_1 + i {\bf N}_2 = \left( \begin{array}{c} z_{\downarrow}^2
- z_{\uparrow}^2 \\
i (z_{\uparrow}^2 + z_{\downarrow}^2 ) \\
 2 z_{\uparrow} z_{\downarrow} \end{array}
\right) \label{zz}
\end{equation}
It can also be checked from (\ref{zz}) that $(z_{\uparrow},
z_{\downarrow})$ transforms like an $S=1/2$ spinor under spin
rotations. So instead of dealing with a constrained theory of
${\bf N}_{1,2}$ fluctuations, we can express the theory in terms
of the complex spinor $(z_\uparrow, z_\downarrow)$, which is free
of constraints. There is one crucial price we have to pay for this
simplification: notice that the parametrization (\ref{zz}) is {\em
double-valued} and that the spinors $(z_\uparrow, z_\downarrow)$
and $(-z_\uparrow, -z_\downarrow)$ both correspond to the same
non-collinearly ordered state. Indeed, we can change the sign of
$z$ independently at different points in spacetime without
changing the physics, and so any effective action for the
$(z_\uparrow, z_\downarrow)$ spinor must be obey a $Z_2$ gauge
invariance. Here is our first connection with the topologically
ordered paramagnetic state of Section~\ref{sec:topo}, where we had
also discussed a description by a $Z_2$ gauge theory.

\begin{figure}
\centerline{\includegraphics[width=3in]{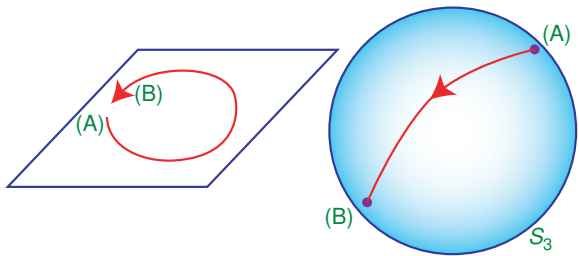}}
\caption{A vison \protect\cite{sf00}. On the left we show a
circular path in real space; this path could be entirely within
the bulk of the system (in which case it defines a local vison
excitation) or it encircles the entire system, which obeys
periodic boundary conditions (so that now it defines a global
topological excitation). On the right is the space of magnetically
ordered states represented by the complex spinor $(z_{\uparrow},
z_{\downarrow})$ up to an overall sign. As we traverse the real
space circle, the path in order parameter space connects polar
opposite points on $S_3$ (A and B), which are physically
indistinguishable. A key point is that this vison excitation can
be defined even in a state in which magnetic order is lost: the
path on the right will fluctuate all over the sphere in quantum
imaginary time, as will the location of the points A and B, but A
and B will remain polar opposites.} \label{fig11}
\end{figure}

In the magnetically ordered non-collinear state we expect dominant
rotational fluctuations about some fixed non-zero value ${\bf
N}_1^2 = {\bf N}_2^2 = (|z_\uparrow |^2 + |z_\downarrow |^2)^2$.
The constraint $|z_\uparrow |^2 + |z_\downarrow |^2 =
\mbox{constant}$ defines the surface of a sphere in a
four-dimensional space ($S_3$) of magnetically ordered ground
states defined by the real and imaginary components of
$z_\uparrow$, $z_\downarrow$. However, we need to identify
opposite points on the sphere with each other, as $(z_\uparrow,
z_\downarrow)$ and $(-z_\uparrow, -z_\downarrow)$ are equivalent
states: this identifies the order parameter space with $S_3 /Z_2$.
This quotient form has crucial consequences for the topological
defect excitations that are permitted in both the magnetically
ordered and the paramagnetic phases. In particular, the order
parameter space (see the review article by \textcite{Mermin79})
allows stable $Z_2$ vortices associated with the first homotopy
group $\pi_1 (S_3 / Z_2 ) = Z_2$: upon encircling such a vortex,
we traverse a path in the order parameter space from $(z_\uparrow
, z_\downarrow)$ to $(-z_\uparrow , -z_\downarrow)$, as shown in
Fig~\ref{fig11}.
As argued in the caption, a fundamental point is that such
vortices can be defined as sensible excitations even in the
paramagnetic phase, where $(z_\uparrow , z_\downarrow )$ is
strongly fluctuating in quantum imaginary time: upon encircling
the vortex, the path in order parameter space will also strongly
fluctuate, but will always connect polar opposite points on $S_3$.
We identify these paramagnetic vortices with the visons of
Section~\ref{sec:topo}, thus firmly establishing a connection
between  non-collinear magnetic order and the topologically
ordered paramagnet.

Finally, we wish to consider a $Z_2$ gauge theory in which
magnetic order is lost continuously \cite{rs91,css94a}, and we
obtain a paramagnetic phase in which the spinor $(z_\uparrow,
z_\downarrow)$ fluctuates about $0$. A pedagogical description of
such a theory was provided by \textcite{lammert93} and
\textcite{lammert95} in an entirely different context: they
considered thermal phase transitions in a nematic liquid crystal,
with order parameter $S_2/Z_2$, in three spatial dimensions.
However their results can be transposed to the quantum phase
transition in two spatial and one imaginary time dimension of
interest here, with the primary change being in the order
parameter space from $S_2/Z_2$ to $S_3/Z_2$: this change is only
expected to modify uninteresting numerical factors in the phase
diagram, as the global topologies of the two spaces are the same.
As shown by \textcite{lammert93} and \textcite{lammert95}, the
magnetically ordered state (with states labeled by points in
$S_{2,3}/Z_2$) does indeed undergo a continuous phase transition
to a paramagnetic state in which spin rotation invariance is
restored and a topological order is present. This topological
order arises because the $Z_2$ visons discussed in Fig~\ref{fig11}
do not proliferate in the paramagnetic state; in this sense, the
topological order here is similar to the topological order in the
low temperature phase of the classical XY model in 2 dimensions,
where point vortices are suppressed below the Kosterlitz-Thouless
transition \cite{thoulessbook}. We can also connect the
nonproliferation of visons to our discussion in
Section~\ref{sec:topo}, where we noted that there was an
excitation gap towards the creation of $Z_2$ visons \cite{sf00}.
Indeed, an explicit connection between the topological order being
discussed here and the topological order noted in the caption to
Fig~\ref{fig10} was established by \textcite{rs91},
\textcite{sr91}, and \textcite{css94a}.

Moreover, without the proliferation of visons in the ground state,
the $(z_\uparrow , z_\downarrow )$ configurations can be defined
as single-valued configurations throughout the sample. Normal-mode
oscillations of $(z_\uparrow , z_\downarrow )$ about zero can now
be identified as a neutral $S=1/2$ particle. This is clearly
related to the spinon excitation of Section~\ref{sec:topo}; this
is our final confirmation of the intimate connection between the
non-collinear magnetic states of Section~\ref{sec:noncoll} and the
topologically ordered states of Section~\ref{sec:topo}.

This is a good point to mention, in passing, recent neutron
scattering evidence for a RVB state in Cs$_2$CuCl$_4$
\cite{coldea01}; the measurements also show non-collinear spin
correlations, consistent with the connections being drawn here.

\section{ORDER IN STATES PROXIMATE TO MOTT INSULATORS}
\label{sec:dope}

We are now ready to discuss the central issue of order parameters
characterizing the cuprate superconductors. These superconductors
are obtained by introducing mobile charge carriers into the Mott
insulator of the square lattice of Cu ions that was discussed at
the beginning of Section~\ref{sec:mott}. The charge carriers are
introduced by substitutional doping. For instance, in the compound
La$_{2-\delta}$Sr$_{\delta}$CuO$_4$, each trivalent La$^{3+}$ ion
replaced by a divalent Sr$^{2+}$ ion causes one hole to appear in
the Mott insulator of Cu ions: the concentration of these holes is
$\delta$ per square lattice site.

For large enough $\delta$, theory and experiment both indicate
that such a doped Mott insulator is a $d$-wave superconductor
characterized by the pairing amplitude (\ref{pair}). The reader
can gain an intuitive (but quite crude and incomplete)
understanding of this by the similarity between the real-space,
short-range pair in (\ref{singlet}) and the momentum-space,
long-range pairing in (\ref{pair}). The undoped Mott insulator
already has electrons paired into singlet valence bonds, as in
(\ref{singlet}), but the repulsive Coulomb energy of the Mott
insulator prevents motion of the charge associated with this pair
of electrons. It should be clear from our discussion in
Section~\ref{sec:para} that this singlet pairing is complete in
the paramagnetic Mott insulators, but we can also expect a partial
pairing in the magnetically ordered states. Upon introducing holes
into the Mott insulator, it becomes possible to move charges
around without any additional Coulomb energy cost, and so the
static valence bond pairs in (\ref{singlet}) transmute into the
mobile Cooper pairs in (\ref{pair}); the condensation of these
pairs leads to superconductivity. Note that this discussion is
concerned with the nature of the ground state wavefunction, and we
are not implying a ``mechanism'' for the formation of Cooper
pairs.

The discussion in the previous sections has laid the groundwork
for a more precise characterization of this superconductor using
the correlations of various order parameters, and of their
interplay with each other. The proximity of the Mott insulator
indicates that the Cooper pairs should be considered descendants
of the real-space, short range pairs in (\ref{singlet}), and this
clearly demands that all the magnetic, bond and topological order
parameters discussed in Section~\ref{sec:mott} remain viable
candidates for the doped Mott insulator. The motion of charge
carriers allows for additional order parameters, and the most
important of these is clearly the superconducting order of the BCS
state noted below (\ref{pair}) in Section~\ref{sec:bcs}. In
principle, it is also possible to obtain new order parameters
which are characteristic of {\em neither} the BCS state {\em nor}
a Mott insulator, but we such order parameters shall not be
discussed here (discussions of one such order may be found in
\textcite{hsumarston91}, \textcite{wenlee96}, \textcite{leesha02},
\textcite{clmn01}, and \textcite{scfmt02}).

The arsenal of order parameters associated with Mott insulators
and the BCS state permits a very wide variety of possible phases
of doped Mott insulators, and of quantum phase transitions between
them. Further progress requires experimental guidance, but we
claim that valuable input is also obtained from the theoretical
connections sketched in Section~\ref{sec:connect}.

The simplest line of reasoning \cite{sr91} uses the fact that the
undoped Mott insulator La$_2$CuO$_4$ has collinear magnetic order
as sketched in Fig~\ref{fig3}a. The arguments above and those in
Section~\ref{sec:connect} then imply that the doped Mott insulator
should be characterized by the collinear magnetic order of
Section~\ref{sec:collinear}, the bond order of
Section~\ref{sec:bond}, along with the phase order of BCS theory.
This still permits a large variety of phase diagrams, and some of
these were explored in \textcite{sr91}, \textcite{vs99},
\textcite{vzs00}, and \textcite{vojta02}, with detailed results on
the evolution of bond order and superconductivity with increasing
doping. However, this reasoning excludes phases associated with
the non-collinear magnetic order of Section~\ref{sec:noncoll} and
the topological order of Section~\ref{sec:topo}.

Some support for this line of reasoning came from the breakthrough
experiments of \textcite{tranquada95}, \textcite{tranquada96}, and
\textcite{tranquada97} on
La$_{2-y-\delta}$Nd$_y$Sr$_{\delta}$O$_4$ for hole concentrations
near $\delta = 1/8$: they observed static, collinear magnetic
order near the wavevectors $\vec{K} = (3 \pi/4, \pi)$ shown in
Figs~\ref{fig3}b,c, which co-existed microscopically\footnote{The
microscopic co-existence of magnetic order and superconductivity
is not universally accepted, but strong arguments in its favor
have been made recently by \protect\textcite{khaykovich02a}.} with
superconductivity for most $\delta$. They also observed
modulations in the bond order $Q_a (\vec{r}_j )$ (Eqn
(\ref{bond})) at the expected wavevector, $2 \vec{K}$. The
experimentalists interpreted their observations in terms of
modulations of the site charge density---proportional to $Q_0
(\vec{r}_j )$---but the existing data actually do not discriminate
between the different possible values of $\vec{r}_a$. As we noted
earlier in Section~\ref{sec:bond}, the physical considerations of
the present article suggest that the modulation may be stronger
with $\vec{r}_a \neq 0$. (The existing data also cannot
distinguish between the magnetic orders in Fig~\ref{fig3}b
(site-centered) and Fig~\ref{fig3}c (bond-centered), or between
the bond orders in Fig~\ref{fig6}c (orthorhombic symmetry) and
Fig~\ref{fig6}d (tetragonal symmetry).) We also mention here the
different physical considerations in the early theoretical work of
\textcite{zaanen89}, \textcite{machida89}, \textcite{schulz89},
and \textcite{poilblanc89} which led to insulating states with
collinear magnetic order with wavevector $\vec{K} \neq (\pi, \pi)$
driven by a large site-charge density modulation in the domain
walls of holes.

The following subsections discuss a number of recent experiments
which explore the interplay between the order parameters we have
introduced here. We argue that all of these experiments support
the proposal that the cuprate superconductors are characterized by
interplay between the collinear magnetic order of
Section~\ref{sec:collinear}, the bond order of
Section~\ref{sec:bond} (these are connected as discussed in
Section~\ref{sec:collbond}), and the superconducting order of BCS
theory.

\subsection{Tuning order by means of a magnetic field}
\label{sec:tune}

In Section~\ref{sec:intro}, we identified a valuable theoretical
tool for the study of systems with multiple order parameters: use
a coupling $g$ to tune the relative weights of static or
fluctuating order parameter correlations in the ground state. Is
such a coupling available experimentally ? One choice is the hole
concentration, $\delta$, and we can assume here that $g$ increases
monotonically with $\delta$. However, $\delta$ is often difficult
to vary continuously, and it may be that sampling the phase
diagram along this one-dimensional axis may not reveal the full
range of physically relevant behavior. A second tuning parameter
will be clearly valuable; here we argue that, under suitable
conditions, this is provided by a magnetic field applied
perpendicular to the two-dimensional layers.

Consider the case where both phases in Fig~\ref{fig1} are
superconducting; the phase with $g<g_c$ then has co-existence of
long-range order in superconductivity and a secondary order
parameter. We also restrict attention to the case where the
transition at $g=g_c$ is second order (related results apply also
to first order transitions, but we do not discuss them here).
Imposing a magnetic field, $H$, on these states will induce an
inhomogeneous state, consisting of a lattice of vortices
surrounded by halos of superflow (we assume here that $H >
H_{c1}$, the lower critical field for flux penetration). In
principle, we now need to study the secondary order parameter in
this inhomogenous background, which can be a problem of some
complexity. However, it was argued by \textcite{dsz01} and
\textcite{zds02} that the problem simplifies considerably near the
phase boundary at $g=g_c$. Because of the diverging correlation
length associated with the secondary order parameter, we need only
look at the spatially-averaged energy associated with the relevant
order parameters.\footnote{Evidence that the primary effect of a
magnetic field is a spatially uniform modification of the magnetic
order has appeared in recent muon spin resonance experiments
\cite{sonier2003,uemura2003}.} We know from the standard theory of
the vortex lattice in a BCS superconductor \cite{parks69} that the
energy density of the superconducting order increases by the
fraction $\sim (H/H_{c2}) \ln (H_{c2}/H)$, where $H_{c2} \gg
H_{c1}$ is the upper critical field above which superconductivity
is destroyed. Let us make the simple assumption that this change
in energy of the superconducting order can also be characterized
by a change in the coupling constant $g$. We can therefore
introduce an effective coupling $g_{\rm eff} (H)$ given by
\begin{equation}
g_{\rm eff} (H)= g - \mathcal{C}' \left( \frac{H}{H_{c2}} \right)
\ln \left( \frac{H_{c2}}{H} \right) \label{geff}
\end{equation}
where $\mathcal{C}'$ is some constant of order unity. As $g$ is
linearly related to $\delta$, we can also rewrite this expression
in terms of an effective doping concentration $\delta_{\rm eff}
(H)$,
\begin{equation}
\delta_{\rm eff} (H)= \delta - \mathcal{C} \left( \frac{H}{H_{c2}}
\right) \ln \left( \frac{H_{c2}}{H} \right), \label{deff}
\end{equation}
where $\mathcal{C}$ is some other constant. These expressions
imply that we tune through different values of $g$ or $\delta$
simply by varying the applied magnetic field.

The sign of $\mathcal{C}$ is of some physical importance, and can
be deduced by a simple argument. It is observed that in the
lightly doped cuprates, decreasing $\delta$ leads to a
stabilization of an order associated with the Mott insulator at
the expense of the superconducting order. (There is a
non-monotonic dependence on $\delta$ from commensurability effects
near $\delta=1/8$, but here too the magnetic order is stabilized
at the expense of superconductivity). As increasing $H$ clearly
suppresses the superconducting order, it must be the case that
$\delta_{\rm eff} (H)$ decreases with increasing $H$. This implies
that $\mathcal{C}>0$, and indicates a competition
\cite{tranquada97} between the two ground states, or order
parameters, on either side of the quantum critical point
\cite{csy,science,so5}.

\begin{figure}
\centerline{\includegraphics[width=3in]{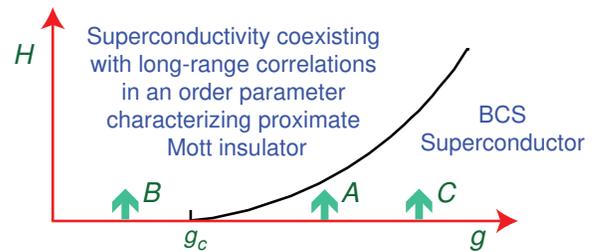}}
\caption{Phase diagram in the $g, H$ plane deduced from
(\protect\ref{geff}). The phase boundary is determined by setting
$g_{\rm eff} (H) = g_c$, which leads to a phase boundary at a
critical field $H \sim (g-g_c)/\ln(1/(g-g_c))$. We assume that $g$
is a monotonically increasing function of $\delta$. The collinear
magnetic order of Figs~\protect\ref{fig3}b and c is the secondary
order parameter investigated in recent neutron scattering
experiments in doped La$_2$CuO$_4$: the observations of
\protect\textcite{lake01} are along the arrow A, and those of
\protect\textcite{katano00}, \protect\textcite{lake02},
\protect\textcite{khaykovich02a}, and
\protect\textcite{khaykovich02b} are along the arrow B. The STM
experiments of \protect\textcite{hoffman02a},
\protect\textcite{hoffman02b}, \protect\textcite{howald02a},
\protect\textcite{howald02b} are along arrow C, and will be
discussed in Section~\protect\ref{sec:stm}.} \label{fig12}
\end{figure}

The relationships (\ref{geff}) and (\ref{deff}) can be combined
with Fig~\ref{fig1} to produce a phase diagram in the $(g, H)$ (or
$(\delta, H)$) plane. This is shown in Fig~\ref{fig12}.
Notice that the phase boundary comes into the $g=g_c$, $H=0$ point
with vanishing slope. This implies that a relatively small field
is needed in the $g>g_c$ region to tune a BCS superconductor
across a quantum phase transition into a state with long-range
correlations in the secondary order parameter. There are also some
interesting modifications to Fig~\ref{fig12} in the fully
three-dimensional model which accounts for the coupling between
adjacent CuO$_2$ layers; these are discussed by
\textcite{kivelson02a}.

A number of neutron scattering studies of the physics of
Fig~\ref{fig12} in doped La$_2$CuO$_4$ have recently appeared. The
secondary order parameter here is the collinear magnetic order of
Figs~\ref{fig3}b and c, which is also observed in
La$_{2-y-\delta}$Nd$_y$Sr$_{\delta}$O$_4$ as discussed above.
Earlier, a series of beautiful experiments by
\textcite{wakimoto99}, \textcite{ylee99}, and
\textcite{wakimoto01} established that
La$_{2-\delta}$Sr$_{\delta}$CuO$_4$ has long-range, collinear
magnetic order co-existing with superconductivity for a range of
$\delta$ values above $\delta = 0.055$. Moreover, the anomalous
frequency and temperature dependence of the dynamic spin structure
factor \cite{sy92,csy} in neutron scattering experiments by
\textcite{aeppli97} gave strong indications of a second-order
quantum phase transition near $\delta \approx 0.14$ at which the
magnetic order vanished. We identify this transition with the
point $g=g_c$, $H=0$ in Fig~\ref{fig12}. Recent studies have
explored the region with $H > 0$: \textcite{lake01} observed a
softening of a collective spin excitation mode at $\delta=0.163$
in the presence of an applied magnetic field. We interpret this as
a consequence of the low $H$ approach to the phase boundary in
Fig~\ref{fig12} in the $g>g_c$ region, as indicated by the arrow
labeled $A$. Notice that the field was not large enough to cross
the phase boundary.

A separate set of experiments have examined the $H$ dependence of
the static magnetic moment in the superconductor with $g< g_c$ in
La$_{2-\delta}$Sr$_{\delta}$CuO$_4$ \cite{katano00,lake02} and
La$_2$CuO$_{4+y}$ \cite{khaykovich02a,khaykovich02b}, along the
arrow indicated by $B$ in Fig~\ref{fig12}. The theoretical
prediction \cite{dsz01,zds02} for these experiments is a simple
consequence of (\ref{geff}) and (\ref{deff}). Let $I(H, \delta)$
be the observed intensity of the static magnetic moment associated
with the order in Figs~\ref{fig3}b,c at a field $H$ and doping
$\delta$. If we assume that the dominant effect of the field can
be absorbed by replacing $\delta$ by the effective $\delta_{\rm
eff} (H)$, we can write
\begin{eqnarray}
I(H, \delta) & \approx & I (H=0, \delta_{\rm eff} (H)) \nonumber
\\
& \approx & I(H=0, \delta) + \mathcal{D} \left( \frac{H}{H_{c2}}
\right) \ln \left( \frac{H_{c2}}{H} \right), \label{ilogh}
\end{eqnarray}
where in the second expression we have used (\ref{deff}) and
expanded in powers of the second argument of $I$. Reasoning as in
the text below (\ref{deff}) for $\mathcal{C}$, we use the
experimental fact that a decrease in $\delta$ leads to an increase
in the magnetic order, and hence $\mathcal{D}
> 0$. The results of recent experiments
\cite{lake02,khaykovich02a,khaykovich02b} are in good agreement
with the prediction (\ref{ilogh}), with a reasonable value for
$\mathcal{D}$ obtained by fitting (\ref{ilogh}) to the
experimental data.

\subsection{Detecting topological order}
\label{sec:senthil}

The magnetic and bond orders break simple symmetries of the
Hamiltonian, and, at least in principle, these can be detected by
measurement of the appropriate two-point correlation function in a
scattering experiment. The topological order of
Sections~\ref{sec:topo} and~\ref{sec:noncolltopo} is a far more
subtle characterization of the electron wavefunction, and can only
be observed indirectly through its consequences for the low energy
excitations. We review here the rationale behind some recent
experimental searches \cite{wynn01,bonn01} for topological order.

The searches relied on a peculiar property of a superconductor
proximate to a Mott insulator with topological order: there is a
fundamental distinction in the internal structure of vortices in
the superconducting order, specified by (\ref{wind}), which
depends on whether the integer $n_v$ is even or odd. This
difference was noted \cite{ss92,nl92} in the context of a simple
mean-field theory of a superconductor near an insulating spin gap
state. However, the significance and interpretation of the
mean-field result, and in particular its connection with
topological order, did not become apparent until the far-reaching
work of \textcite{sf00}, \textcite{sf01a}, and \textcite{sf01b}.
The arguments behind the dependence on the parity of $n_v$ are
subtle, and only an outline will be sketched here---the reader is
referred to \textcite{sf01a} and \textcite{sf01b} for a complete
exposition. Although the superconducting order of BCS theory in
(\ref{pair}) and the topological order of the Mott insulator are
quite distinct entities, there is an important connection between
them in the superconducting state: each vortex with $n_v$ odd in
(\ref{wind}) has a vison attached to it. The vison gap in the
proximate Mott insulator then increases the energy required to
create $n_v$ odd vortices, while this extra energy is not required
for $n_v$ even.

The connection between $n_v$ odd vortices and visons is most
transparent for the case where the spinons in the Mott insulator
obey fermionic statistics. We considered bosonic spinons
$z_{\sigma}$ in Section~\ref{sec:noncolltopo}, but they can
transmute into fermions by binding with a vison
\cite{rc89,kivelson89,demler02}: we represent the fermionic spinon
by $f_{j \sigma}$. In the doped Mott insulator, each electron
annihilation operator, $c_{j \sigma}$, must create at least one
neutral $S=1/2$ spinon excitation, along with a charge $e$ hole
\cite{krs87}, and we can represent this schematically by the
operator relation
\begin{equation}
c_{j \sigma} = b^{\dagger}_j f_{j \sigma}, \label{spincharge}
\end{equation}
where $b_j^{\dagger}$ creates a bosonic spinless hole. In this
picture of the doped Mott insulator, the presence of
superconductivity as in (\ref{pair}) requires both the
condensation of the $b_j$, along with the condensation of ``Cooper
pairs'' of the spinons $f_{j \sigma}$. We can deduce this
relationship from (\ref{pair}) and (\ref{spincharge}) which imply,
schematically
\begin{equation}
\Delta_0 = \Delta_f b^2, \label{dfb}
\end{equation}
where we have ignored spatial dependence associated with the
internal wavefunction of the Cooper pair (hence there are no site
subscripts $j$ in (\ref{dfb})), and $\Delta_f \sim \langle f_{j
\uparrow} f_{j' \downarrow} \rangle$ is the spinon pairing
amplitude. From (\ref{dfb}) we see if the phase of $b_j$ winds by
$2 \pi$ upon encircling some defect site, then phase of $\Delta_0$
will wind by $4 \pi$, and this corresponds to a vortex in the
superconducting order with $n_v=2$ in (\ref{wind}). Indeed, the
only way (\ref{dfb}) can lead to an elementary vortex with $n_v =
1$ is if the phase of the spinon pair amplitude, $\Delta_f$, winds
by $2 \pi$ upon encircling the vortex: the latter is another
description of a vison \cite{sf00}. This argument is easily
extended to show that every odd $n_v$ vortex must be associated
with at least an elementary vortex in the phase of $\Delta_f$,
thus establishing our claimed connection.

Sufficiently close to the Mott insulator, and near a second-order
superconductor-insulator transition, the energy required to create
a vison raises the energy of $n_v = 1 $ vortices, and the lowest
energy vortex lattice state in an applied magnetic field turns out
to have vortices with flux $hc/e$, which is twice the elementary
flux \cite{ss92}. This should be easily detectable, but such
searches have not been successful so far \cite{wynn01}.

More recently \textcite{sf01a} have proposed an ingenious test for
the presence of visons, also relying on the binding of a vison to
a vortex with fluc $hc/(2e)$. Begin with a superconductor in a
toroidal geometry with flux $hc/(2e)$ penetrating the hole of the
torus. By the arguments above, a vison is also trapped in the hole
of the torus. Now by changing either the temperature or the doping
level of the superconductor, drive it into a normal state. This
will allow the magnetic flux to escape, but the topological order
in the bulk will continue to trap the vison. Finally, return the
system back to its superconducting state, and, quite remarkably,
the vison will cause the magnetic flux to reappear. An
experimental test for this ``flux memory effect'' has also been
undertaken \cite{bonn01}, but no such effect has yet been found.

So despite some innovative and valuable experimental tests, no
topological order has been detected so far in the cuprate
superconductors.

\subsection{Non magnetic impurities}
\label{sec:imp}

We noted in Section~\ref{sec:bond} that one of the key
consequences of the confinement of spinons in the bond-ordered
paramagnet was that each non-magnetic impurity would bind a free
$S=1/2$ moment. In contrast, in the topologically ordered RVB
states of Section~\ref{sec:topo}, such a moment is not generically
expected, and it is more likely that the ``liquid'' of valence
bonds would readjust itself to screen away the offending impurity
without releasing any free spins.

Moving to the doped Mott insulator, we then expect no free $S=1/2$
moment for the topologically ordered case. The remaining
discussion here is for the confining case; in this situation the
$S=1/2$ moment may well survive over a finite range of doping,
beyond that required for the onset of superconductivity.
Eventually, at large enough hole concentrations, the low energy
fermionic excitations in the $d$-wave superconductor will screen
the moment (by the Kondo effect) at the lowest temperatures.
However, unlike the case of a Fermi liquid, the linearly vanishing
density of fermionic states at the Fermi level implies that the
Kondo temperature can be strictly zero for a finite range of
parameters \cite{withoff90,ingersent98,vojtabulla02}. So we expect
each non-magnetic impurity to create a free $S=1/2$ moment that
survives down to $T=0$ for a finite range of doping in a $d$-wave
superconductor proximate to a confining Mott insulator. The
collinear magnetic or bond order in the latter insulator may also
survive into the superconducting state, but there is no
fundamental reason for the disappearance of these long-range
orders (bulk quantum phase transitions) to coincide with the zero
temperature quenching of the moment (an impurity quantum phase
transition).

A very large number of experimental studies of non-magnetic Zn and
Li impurities have been carried out. Early on, in electron
paramagnetic resonance experiments \textcite{fink90} observed the
trapping of an $S=1/2$ moment near a Zn impurity above the
superconducting critical temperature, and also noted the
implication of their observations for the confinement of spinons,
in the spirit of our discussion above. Subsequent specific heat
and nuclear magnetic resonance experiments
\cite{alloul91,sisson00,julien00,bobroff01} have also explored low
temperatures in the superconducting state, and find evidence of
spin moments, which are eventually quenched by the Kondo effect in
the large doping regime. Especially notable is the recent nuclear
magnetic resonance evidence \cite{bobroff01} for a transition from
a $T=0$ free moment state at low doping, to a Kondo quenched state
at high doping.

We interpret these results as strong evidence for the presence of
an $S=1/2$ moment near non-magnetic impurities in the lightly
doped cuprates. We have also argued here, and elsewhere
\cite{icmp}, that the physics of this moment formation is most
naturally understood in terms of the physics of a proximate Mott
insulator with spinon confinement.

The creation of a free magnetic moment (with a local magnetic
susceptibility which diverges as $\sim 1/T$ as $T \rightarrow 0$)
near a single impurity implies that the cuprate superconductors
are exceptionally sensitive to disorder. Other defects, such as
vacancies, dislocations, and grain boundaries, which are
invariably present even in the best crystals, should also have
similar strong effects. We speculate that it is this tendency to
produce free moments (and local spin order which will be induced
in their vicinity) which is responsible for the frequent recent
observation of magnetic moments in the lightly doped cuprates
\cite{sonier01,sidis01}.

\subsection{STM studies of the vortex lattice}
\label{sec:stm}

Section~\ref{sec:tune} discussed the tuning of collinear magnetic
order by means of an applied magnetic field, and its detection in
neutron scattering experiments in doped La$_2$CuO$_4$. This
naturally raises the question of whether it may also be possible
to detect the bond order of Section~\ref{sec:bond} somewhere in
the phase diagram of Fig~\ref{fig12}. Clearly the state with
co-existing collinear magnetic and superconducting order (explored
by experiments along the arrow B) should, by the arguments of
Section~\ref{sec:collbond}, also have co-existing bond order.
However, more interesting is the possibility that the BCS
superconductor itself has local regions of bond order for $H \neq
0$ \cite{park01}. As we have argued, increasing $H$ increases the
weight of the Mott insulator order parameter correlations in the
superconducting ground state. The appearance of static magnetic
order requires breaking of spin rotation invariance (in the plane
perpendicular to the applied field), and this cannot happen until
there is a bulk phase transition indicated by the phase boundary
in Fig~\ref{fig12}. In contrast, bond order only breaks
translational symmetry, but this is already broken by the vortex
lattice induced by a non-zero $H$. The small vortex cores can pin
the translational degree of freedom of the bond order, and a halo
of static bond order should appear around each vortex core
\cite{park01,psvd01,dsz01,zds02,pvs02}. Notice that this bond
order has appeared in the state which has only superconducting
order at $H=0$, and so should be visible along the arrow labelled
C in Fig~\ref{fig12}. Recall also our discussion in
Section~\ref{sec:bond} that site charge order is a special case of
bond order (with $\vec{r}_a = 0$ in the bond order parameter $Q_a
(\vec{r})$).

Many other proposals have also been made for additional order
parameters within the vortex core. The earliest of these
 involved dynamic antiferromagnetism\cite{ss92,nl92}, and were discussed in
Section~\ref{sec:senthil} in the context of topological order.
Others
\cite{so5,arovas97,chen02,zhu02,franz02,chenhu02,ghosal02,andersen02,ichioka02}
involve static magnetism within each vortex core in the
superconductor.\footnote{Also, \protect\textcite{ichioka01}
studied the vortex lattice in a state with pre-existing long-range
collinear spin order} This appears unlikely from the perspective
of the physics of Fig~\ref{fig12}, in which static magnetism only
appears after there is a co-operative bulk transition to
long-range magnetic order, in the region above the phase boundary;
below the phase boundary there are no static ``spins in
vortices,'' but there is bond order as discussed above
\cite{park01,zds02}. (Static spins do appear in the three space
dimensional model with spin anisotropy and inter-planar couplings
considered in \textcite{kivelson02a}.) A separate proposal
involving staggered current loops in the vortex core
\cite{kishine01,leesha02,lee01} has also been made.

Nanoscale studies looking for signals of bond order along the
arrow C in Fig~\ref{fig12} would clearly be helpful. Scanning
tunnelling microscopy (STM) is the ideal tool, but requires
atomically clean surfaces of the cuprate crystal. The detection of
collinear magnetic order in doped La$_2$CuO$_4$ makes such
materials ideal candidates for bond order, but they have not been
amenable to STM studies so far. Crystals of
Bi$_2$Sr$_2$CaCu$_2$O$_{8+\delta}$ have been the focus of numerous
STM studies, but there is little indication of magnetic order in
neutron scattering studies of this superconductor. Nevertheless,
by the reasoning in Fig~\ref{fig12}, and using the reasonable
hypothesis that a common picture of competing superconducting,
bond, and collinear magnetic order applies to all the cuprates, it
is plausible that static bond order should appear in
Bi$_2$Sr$_2$CaCu$_2$O$_{8+\delta}$ for large enough $H$ along the
arrow C in Fig~\ref{fig12}.

A number of atomic resolution STM studies of
Bi$_2$Sr$_2$CaCu$_2$O$_{8+\delta}$ surfaces have appeared recently
\cite{hoffman02a,hoffman02b,howald02a,howald02b}.
\textcite{hoffman02a} observed a clear signal of modulations in
the local density of electronic states, with a period of 4 lattice
spacings, in a halo around each vortex core. There was no
corresponding modulation in the surface topography, implying there
is little modulation in the charge density. However, a bond order
modulation, such as those in Figs~\ref{fig6}c and d, could
naturally lead to the required modulation in the local density of
states. Other studies \cite{hoffman02b,howald02a,howald02b} have
focused on the $H=0$ region: here the modulations appear to have
significant contributions \cite{byers93,wang02} from scattering of
the fermionic $S=1/2$ quasiparticles of the superconductor
(Section~\ref{sec:bcs}), but there are also signals
\cite{howald02a,howald02b} of a weak residual periodic modulation
in the density of states, similar to those found at $H\neq 0$.
Theoretically \cite{psv02,howald02b,kivelson02}, it is quite
natural that these quasiparticle and order parameter modulations
co-exist. \textcite{howald02a} and \textcite{howald02b} also
presented results for the energy dependence of this periodic
modulation, and these appear to be best modelled by modulations in
microscopic bond, rather than site, variables
\cite{podolsky02,vojta02,zhang02}.

This is a rapidly evolving field of investigation, and future
experiments should help settle the interpretation of the density
of states modulations both at $H=0$ and $H \neq 0$. It should be
noted that because translational symmetry is broken by the
vortices or the pinning centers, there is no fundamental symmetry
distinction between the quasiparticle and the
pinned-fluctuating-order contributions; nevertheless, their
separate spectral and spatial features should allow us to
distinguish them.

\section{A PHASE DIAGRAM WITH COLLINEAR SPINS, BOND ORDER,
AND SUPERCONDUCTIVITY} \label{sec:global}

We have already discussed two experimental possibilities for the
coupling $g$ in Fig~\ref{fig1}, which we used to tune the ground
state of the doped Mott insulator between various distinct phases:
the doping concentration, $\delta$, and the strength of a magnetic
field, $H$, applied perpendicular to the layers. A simple phase
diagram in the small $H$ region as a function of these parameters
was presented in Fig~\ref{fig12}, and its implications were
compared with a number of experiments in Sections~\ref{sec:tune}
and~\ref{sec:stm}. However, even though it is experimentally
accessible, the field $H$ induces a large scale spatial modulation
associated with the vortex lattice, and is consequently an
inconvenient choice for microscopic theoretical calculations. Here
we follow the strategy of introducing a third theoretical axis,
which we denote schematically by $\widetilde{g}$, to obtain a
global view of the phase diagram. As we argue below, information
on the phases present as a function of $\widetilde{g}$ sheds
considerable light on the physics as a function of $H$.

The crucial role of order parameters characterizing Mott
insulators in our discussion suggests that we should work with a
coupling, $\widetilde{g}$, which allows exploration of different
ground states of Mott insulators already at $\delta = 0$. The
range of this coupling should obviously include regimes where the
Mott insulator has the magnetically ordered ground state of
Fig~\ref{fig3}a, found in La$_2$CuO$_4$. Now imagine adding
further neighbor couplings in (\ref{afm}) which frustrate this
magnetic order, and eventually lead to a phase transition to a
paramagnetic state.\footnote{We assume that there is no
intermediate state with non-collinear magnetic order, as this is
not supported by observations so far.} As discussed in
Section~\ref{sec:collbond}, it was argued \cite{rs89b,rs90,sp02}
that any paramagnetic state so obtained should have bond order,
most likely in the patterns in Figs~\ref{fig6}a and b.

It would clearly be useful to have numerical studies which tune a
coupling $\widetilde{g}$ acting in the manner described above.
{\em Large scale} computer studies of this type have only appeared
recently. The first results on a quantum antiferromagnet which has
a spin of $S=1/2$ per unit cell, whose Hamiltonian maintains full
square lattice symmetry, and in which it is possible to tune a
coupling to destroy the collinear magnetic order, were obtained
recently by \textcite{sandvik02}. Their model extended (\ref{afm})
with a plaquette ring-exchange term, and had only a U(1) spin
rotation symmetry. Theoretical extensions to this case have also
been discussed \cite{lannert01,ps02}. Along with the collinear
magnetic state in the small ring-exchange region (small
$\widetilde{g}$), \textcite{sandvik02} found the bond-ordered
paramagnetic state of Fig~\ref{fig6}a in the large ring-exchange
region (large $\widetilde{g}$).

A second large scale computer study of the destruction of
collinear magnetic order on a model with $S=1/2$ per unit cell was
performed recently by \textcite{harada02}. They generalized the
spin symmetry group from SU(2) to SU($N$); in our language, they
used the value of $N$ as an effective $\widetilde{g}$. They also
found the bond order of Fig~\ref{fig6}a in the paramagnetic
region.

\begin{figure}
\centerline{\includegraphics[width=3in]{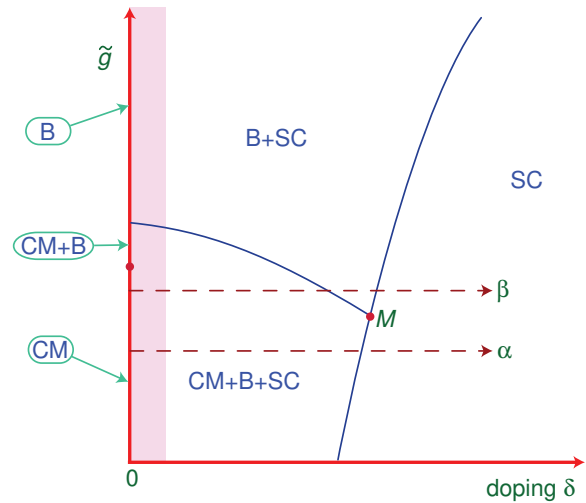}}
\caption{Zero temperature, zero magnetic field phase diagram as a
function of the doping $\delta$, and a coupling constant
$\widetilde{g}$. Here $\widetilde{g}$ is, in principle, any
coupling which can destroy the collinear magnetic order at $(\pi,
\pi)$ in the undoped insulator, while the Hamiltonian maintains
full square lattice symmetry with spin $S=1/2$ per unit cell. The
states are labeled by the orders which exhibit long-range
correlations: collinear magnetic (CM), bond (B) and $d$-wave-like
superconductivity (SC). At $\delta = 0$, the CM order is as in
Fig~\protect\ref{fig3}a, the B order is as in
Fig~\protect\ref{fig6}a or b, and we have assumed a co-existing
CM+B region, following \protect\textcite{sp02} and
\protect\textcite{sushkov01}. The ground state will remain an
insulator for a small range of $\delta
>0$ (induced by the long-range Coulomb interactions), and this
is represented by the shaded region. The CM order for $\delta>0$
could be as in Fig~\protect\ref{fig3}b or c, and the B order as in
Figs~\protect\ref{fig6}a, b, c, or d, but a variety of other
periods are also possible \protect\cite{vs99,vojta02}. The dashed
line $\alpha$ indicates the path followed in
Fig~\protect\ref{fig12} at $H=0$, but the physical situation could
also lie along line $\beta$. A number of other complex phases are
possible in the vicinity of the multicritical point $M$; these are
not shown but are discussed in \protect\textcite{zds02},
\protect\textcite{zaanen02}, and \protect\textcite{zaanen02a} and
also, briefly, in Section~\protect\ref{sec:conc}.} \label{fig13}
\end{figure}

These theoretical studies give us confidence in the theoretical
phase diagram as a function of $\widetilde{g}$ and $\delta$
sketched in Fig~\ref{fig13} \cite{sr91,vs99,vzs00,vojta02}.
Phase diagrams with related physical ingredients, but with
significant differences, appear in the work of \textcite{kfe98}
and \textcite{zaanen99}.

Important input in sketching Fig~\ref{fig13} was provided by
theoretical studies of the effects of doping the bond-ordered
paramagnetic Mott insulator at large $\widetilde{g}$. In this
region without magnetic order, it was argued that a systematic and
controlled study of the doped system was provided by a
generalization of the SU(2) spin symmetry\footnote{The group SU(2)
is identical to the symplectic group Sp(2), but the group SU(2$N$)
is distinct from Sp(2$N$) for $N>1$. Consequently, distinct $1/N$
expansions are generated by  models with SU(2$N$) or Sp(2$N$)
symmetry. The Sp(2$N$) choice better captures the physics
discussed in this article, for reasons explained in
\protect\textcite{sr91}} to Sp(2$N$), followed by an expansion in
$1/N$. This approach directly gives \cite{sr91} a stable
bond-ordered state at $\delta=0$, a stable $d$-wave superconductor
at large $\delta$, and a region in which these two orders co-exist
at small values of $\delta$; all of these phases are nicely in
accord with the overall philosophy of the present article. This
analysis of a model with purely short-range interactions also
found a phase separation instability at small values of $\delta$
\cite{sr91}, whose importance had been emphasized by others
\cite{ekl90,bang91} on different grounds. With long-range Coulomb
interactions no macroscopic phase separation is possible, and we
have to deal with the physics of frustrated phase separation
\cite{ekl90}. The interplay between bond order and $d$-wave
superconductivity has been studied in some detail in this region
\cite{vs99,vzs00,vojta02}: more complex bond ordered structures
with large periods can appear, usually co-existing with
superconductivity (as sketched in Fig~\ref{fig13}). Predictions
were made for the evolution of the ordering wavevector with
$\delta$, and the period 4 structures in Figs~\ref{fig6}c and d
were found to be especially stable over a wide regime of doping
and parameter space.

The phase diagram of Fig~\ref{fig13} also includes a region at
small $\widetilde{g}$, with collinear magnetic order, which is not
directly covered by the above computations. ``Stripe physics''
\cite{zaanen89,schulz89,machida89,poilblanc89}---the accumulation
of holes on sites which are anti-phase domain walls between
N\'{e}el ordered regions---is associated with this region.
However, these stripe analyses treat the magnetic order in a
static, classical manner, and this misses the physics of valence
bond formation that has been emphasized in our discussion here. A
related feature is that their domain walls are fully populated
with holes and are insulating. Upon including quantum fluctuations
accounting for valence bonds, it appears likely to us that the
stripes will have partial filling \cite{kivelson96,nayak97},
acquire bond order, and co-exist with superconductivity, as has
been assumed in our phase diagram in Fig~\ref{fig13}. Indeed, as
we have emphasized throughout, it may well be that the modulation
in the site charge density---which is proportional to $Q_a
(\vec{r})$ with $\vec{r}_a = 0$ in (\ref{bond})---is quite small,
and most of the modulation is for $\vec{r}_a \neq 0$.

The reader should now be able to use the perspective of the phase
diagram in Fig~\ref{fig13} to illuminate our discussion of
experiments in Section~\ref{sec:dope}. The phase diagram in
Fig~\ref{fig12}, used to analyze neutron scattering experiments in
Section~\ref{sec:tune} and STM experiments in
Section~\ref{sec:stm}, has its horizontal axis along the line
labeled $\alpha$ in Fig~\ref{fig13}; the phases that appear in
Fig~\ref{fig12} as a function of increasing $H$ should be related
to those in Fig~\ref{fig13} as a function of increasing
$\widetilde{g}$, although the detailed location of the phase
boundaries is surely different.\footnote{More precisely,
generalizing the arguments leading to (\protect\ref{geff}) and
(\protect\ref{deff}), we can state that the system is
characterized by an effective $\widetilde{g}$ which increases
linearly with $H \ln (1/H)$, and an effective $\delta$ which
decreases linearly with $H \ln (1/H)$} The absence of topological
order in the experiments discussed in Section~\ref{sec:senthil},
is seen in Fig~\ref{fig13} to be related to the absence of states
with non-collinear spin correlations or topological order. The
formation of $S=1/2$ moments near non-magnetic impurities is
understood by the proximity of confining, bond-ordered phases in
Fig~\ref{fig13}. The possible signals of bond order in a
superconductor at $H=0$ in the STM observations of Howald {\em et
al.} \cite{howald02a,howald02b}, may be related to the B+SC phase
along the line $\beta$ in Fig~\ref{fig13}; similarly, the
observations of Hoffman {\em et al.} \cite{hoffman02a} at $H \neq
0$ can be interpreted by the proximity of the B+SC phase at $H=0$.

\section{OUTLOOK}
\label{sec:conc}

The main contention of this article is that cuprate
superconductors are best understood in the context of a phase
diagram containing states characterized by the pairing order of
BCS theory, along with orders associated with Mott insulators; the
evidence so far supports the class of Mott insulators with
collinear spins and bond order.
The interplay of these orders permits a
rich variety of distinct phases, and the quantum critical points
between them offer fertile ground for developing a controlled
theory for intermediate regimes characterized by multiple
competing orders. This approach has been used to analyze and
predict the results of a number of recent neutron scattering,
fluxoid detection, NMR, and STM experiments, as we have discussed
in Sections~\ref{sec:tune}, \ref{sec:senthil}, \ref{sec:imp}, and
\ref{sec:stm}. Further experimental tests have also been proposed,
and there are bright prospects for a more detailed, and ultimately
quantitative, confrontation between theory and experiment.

All of the experimental comparisons here have been restricted to
very low temperatures. The theory of crossovers near quantum
critical points also implies interesting anomalous dynamic
properties at finite temperature \cite{sy92,book}, but these have
not been discussed. However, we did note in Section~\ref{sec:tune}
that the transition involving loss of magnetic order in a
background of superconductivity was a natural candidate for
explaining the singular temperature and frequency dependence
observed in the neutron scattering at $\delta \approx 0.14$.
\cite{aeppli97}

There have also been several recent experimental proposals for a
quantum critical point in the cuprates at $\delta \approx 0.19$,
linked to anomalous quasiparticle damping \cite{valla99},
thermodynamic \cite{tallon01}, or magnetic
\cite{panagopoulos02,panagopoulos02a} properties. The study of
Panagopoulous and collaborators presents evidence for a spin glass
state below the critical doping, and this is expected in the
presence of disorder at dopings lower than that of the point M in
Fig~\ref{fig13}.

Among theoretical proposals, a candidate for a quantum critical
point\cite{zds02,sm02,zaanen02} at large dopings is a novel
topological transition which can occur even in systems with
collinear spin correlations. While the topological order present
in systems with non-collinear spin correlation leads to
fractionalization of the electron (as discussed in
Section~\ref{sec:senthil}), the collinear spin case exhibits a
very different and much less disruptive transition in which the
electron retains its integrity, but the spin and charge collective
modes fractionalize into independent entities. Note that this
fractionalization transition was not explicitly shown in
Fig~\ref{fig13}, and is associated with an additional intermediate
state which may appear near the point $M$. Other theoretical
proposals for quantum critical points are linked to the
bond/charge order\cite{kfe98,seibold98} in Fig~\ref{fig13}, to
order associated with circulating current loops
\cite{clmn01,varma} which has not been discussed in this paper,
and to a time-reversal symmetry breaking
transition\cite{laughlin,vzs00a,kp01,sangiovanni01} between
$d_{x^2 - y^2}$ and $d_{x^2 - y^2} + i d_{xy}$ superconductors.
This last proposal offers a possible explanation of the
quasiparticle damping measurements \cite{valla99}. Note that this
transition does not involve any order associated with the Mott
insulator. Indeed, the $d_{x^2 - y^2} + i d_{xy}$ order can be
understood entirely within the framework of BCS theory, and
experimental support for $d_{x^2 - y^2} + i d_{xy}$
superconductivity in recent tunnelling experiments \cite{dagan01}
appears in the overdoped regime, well away from the Mott
insulator.

\section*{Acknowledgments}

I have benefited from discussions and collaborations with many
physicists: here I would like to especially thank Gabriel Aeppli,
Henri Alloul, Robert Birgeneau, Andrey Chubukov, Seamus Davis,
Eugene Demler, Matthew Fisher, Aharon Kapitulnik, Steve Kivelson,
Christos Panagopoulos, Kwon Park, Anatoli Polkovnikov, T.~Senthil,
Matthias Vojta, Jan Zaanen, and Ying Zhang for valuable
interactions in recent years. This article is based on the
F.~A.~Matsen Endowed Regents Lecture on the Theories of Matter at
the University of Texas at Austin, October 2002. This research was
supported by US NSF Grant DMR 0098226.

\bibliographystyle{apsrmp}

\bibliography{rmp}

\end{document}